\documentclass[aps,prd,twocolumn,showpacs,amsmath,amssymb]{revtex4-1}
\usepackage{amsmath}
\usepackage{graphicx}
\usepackage{subfigure}
\usepackage{epstopdf}
\usepackage{color}
\usepackage{multirow}
\usepackage{setspace}
\usepackage{overpic}
\usepackage{amssymb}
\usepackage[bookmarksnumbered, pdfstartview=FitH,colorlinks,urlcolor=blue, citecolor=blue,linkcolor=blue] {hyperref}
\usepackage{lineno}
\usepackage{bm}
\usepackage{rotating}
\usepackage[utf8]{inputenc}

\let\oldequation\equation
\let\oldendequation\endequation

\renewenvironment{equation}
  {\linenomathNonumbers\oldequation}
  {\oldendequation\endlinenomath}

\begin{document}

%\linenumbers

\title{\boldmath Precision Measurements of $D_s^+ \to \eta  e^+ \nu_e$ and $D_s^+ \to \eta^\prime  e^+ \nu_e$}

%% Saved at => 2022-12-12
\author{
M.~Ablikim$^{1}$, M.~N.~Achasov$^{13,b}$, P.~Adlarson$^{75}$, R.~Aliberti$^{36}$, A.~Amoroso$^{74A,74C}$, M.~R.~An$^{40}$, Q.~An$^{71,58}$, Y.~Bai$^{57}$, O.~Bakina$^{37}$, I.~Balossino$^{30A}$
% ...
\\
\vspace{0.2cm}
(BESIII Collaboration)\\
\vspace{0.2cm} {\it
$^{1}$ Institute of High Energy Physics, Beijing 100049, People's Republic of China\\
% ...
$^{13}$ G.I. Budker Institute of Nuclear Physics SB RAS (BINP), Novosibirsk 630090, Russia\\
% ...
$^{30}$ INFN Sezione di Ferrara, (A)INFN Sezione di Ferrara, I-44122, Ferrara, Italy; (B)University of Ferrara,  I-44122, Ferrara, Italy\\
% ...
$^{36}$ Johannes Gutenberg University of Mainz, Johann-Joachim-Becher-Weg 45, D-55099 Mainz, Germany\\
$^{37}$ Joint Institute for Nuclear Research, 141980 Dubna, Moscow region, Russia\\
% ...
$^{40}$ Liaoning Normal University, Dalian 116029, People's Republic of China\\
% ...
$^{57}$ Southeast University, Nanjing 211100, People's Republic of China\\
$^{58}$ State Key Laboratory of Particle Detection and Electronics, Beijing 100049, Hefei 230026, People's Republic of China\\
% ...
$^{71}$ University of Science and Technology of China, Hefei 230026, People's Republic of China\\
% ...
$^{74}$ University of Turin and INFN, (A)University of Turin, I-10125, Turin, Italy; (B)University of Eastern Piedmont, I-15121, Alessandria, Italy; (C)INFN, I-10125, Turin, Italy\\
$^{75}$ Uppsala University, Box 516, SE-75120 Uppsala, Sweden\\
% ...
$^{b}$ Also at the Novosibirsk State University, Novosibirsk, 630090, Russia\\
% ...
}
}
%% ends here %%

\begin{abstract}
Precision measurements of the semileptonic decays $D_s^+ \to \eta  e^+ \nu_e$ and $D_s^+ \to \eta^\prime  e^+ \nu_e$ are performed with 7.33\,fb$^{-1}$  of $e^+e^-$ collision data collected at center-of-mass energies between 4.128 and 4.226 GeV with the BESIII detector.  
The branching fractions obtained are $\mathcal{B}(D_s^+ \to \eta e^{+} \nu_e)$ = $(2.255\pm0.039_{\rm stat}\pm 0.051_{\rm syst})\%$ and
$\mathcal{B}(D_s^+ \to \eta^{\prime} e^{+} \nu_e)$ = $(0.810\pm0.038_{\rm stat}\pm 0.024_{\rm syst})\%$. Combining these results
with the $\mathcal{B}(D^+\to\eta e^+ \nu_e)$ and $\mathcal{B}(D^+\to\eta^\prime e^+ \nu_e)$ obtained from previous BESIII measurements, the
$\eta-\eta^\prime$ mixing angle in the quark flavor basis is determined to be
$\phi_{\rm P} = (40.0\pm2.0_{\rm stat}\pm0.6_{\rm syst})^\circ$.
Moreover, from the fits to the partial decay rates of
$D_s^+ \to \eta  e^+ \nu_e$ and $D_s^+ \to \eta^\prime  e^+ \nu_e$, the products of the hadronic transition form factors
$f_+^{\eta^{(\prime)}}(0)$ and the modulus of the $c\to s$ Cabibbo-Kobayashi-Maskawa matrix element $|V_{cs}|$ are determined by using different hadronic transition form factor parametrizations. Based on the two-parameter series expansion, the products 
$f^\eta_+(0)|V_{cs}| = 0.4519\pm0.0071_{\rm stat}\pm0.0065_{\rm syst}$ and $f^{\eta^\prime}_+(0)|V_{cs}| = 0.525\pm0.024_{\rm stat}\pm0.009_{\rm syst}$ are extracted. All results determined in this work supersede those measured in the previous BESIII analyses based on the 3.19 fb$^{-1}$ subsample of data at 4.178 GeV.  
\end{abstract}

\maketitle

\oddsidemargin  -0.2cm
\evensidemargin -0.2cm

\section{Introduction}

Experimental studies of the semileptonic decays of charmed mesons are important inputs to further understanding of the weak and strong 
interactions in the charm sector~\cite{Ke:2023qzc}. 
By analyzing their decay dynamics, one can extract the product of the modulus of the Cabibbo-Kobayashi-Maskawa (CKM) matrix
element $|V_{cs(d)}|$ and the hadronic transition form factor, offering insights into charm physics.  
Taking $D^+_s\to \eta^{(\prime)} e^+\nu_e$ as an example, the hadronic transition form factors at zero-momentum transfer $f^{\eta^{(\prime)}}_+(0)$~\cite{Bali:2014pva,Hu:2021zmy,Offen:2013nma,Azizi:2010zj,Duplancic:2015zna,Verma:2011yw,Wei:2009nc,Melikhov:2000yu,Soni:2018adu,Ivanov:2019nqd,Colangelo:2001cv} can be calculated via several theoretical approaches, e.g.,  lattice quantum
chromodynamics~(LQCD)~\cite{Bali:2014pva}, QCD light-cone sum rules
(LCSR)~\cite{Hu:2021zmy,Offen:2013nma,Duplancic:2015zna,Azizi:2010zj}, covariant light-front quark model (LFQM)\cite{Verma:2011yw,Wei:2009nc}, constituent quark model (CQM)~\cite{Melikhov:2000yu}, covariant confined quark model (CCQM) \cite{Soni:2018adu,Ivanov:2019nqd}, and QCD sum rules (QCDSR) \cite{Colangelo:2001cv}. The predicted values for $f^\eta_+(0)$  and $f^{\eta^{\prime}}_+(0)$ are summarized in Table~\ref{tab:FF}. 
Using the value of $|V_{cs}|$ provided by the CKMFitter group~\cite{PDG2022}, the hadronic transition form factors can be extracted, resulting in a stringent test of the theoretical predictions. Alternatively, assuming a $f_+^{\eta^{(\prime)}}(0)$ value predicted by theory leads to $|V_{cs}|$, which is important for test of CKM matrix unitarity.  

\begin{table}[htp]
\centering
\caption{\label{tab:FF}
\small   Theoretical predictions of the hadronic transition form factors at zero-momentum transfer $f^{\eta^{(\prime)}}_+(0)$.}
\begin{tabular}{lcc} \hline
&$f_+^{\eta}(0)$&$f_+^{\eta^\prime}(0)$\\ 
\hline
LQCD(I)~\cite{Bali:2014pva}&$0.542\pm0.013$&$0.404\pm0.025$       \\
LQCD(II)~\cite{Bali:2014pva}&$0.564\pm0.011$&$0.437\pm0.018$      \\
LCSR~\cite{Hu:2021zmy}    &$0.476^{+0.040}_{-0.036}$&$0.544^{+0.046}_{-0.042}$\\
LCSR~\cite{Duplancic:2015zna}    &$0.495^{+0.030}_{-0.029}$&$0.558^{+0.047}_{-0.045}$\\
LCSR~\cite{Offen:2013nma}    &$0.432\pm0.033$&$0.520\pm0.080$\\
LCSR~\cite{Azizi:2010zj}    &$0.45\pm0.14$&$0.55\pm0.18$    \\
LFQM(I)~\cite{Wei:2009nc} &0.50         &0.62            \\
LFQM(II)~\cite{Wei:2009nc}&0.48         &0.60              \\
LFQM~\cite{Verma:2011yw}    &0.76         & --            \\
CQM \cite{Melikhov:2000yu}     &0.78         &0.78             \\
CCQM\cite{Ivanov:2019nqd}  &$0.49\pm0.07$&$0.59\pm0.09$\\ 
CCQM\cite{Soni:2018adu}     &$0.78\pm0.12$&$0.73\pm0.11$\\
QCDSR~\cite{Colangelo:2001cv}    &$0.50\pm0.04$& --           \\
\hline
\end{tabular}
\end{table}

In addition, the $\eta-\eta^\prime$ mixing angle in the quark flavor basis, $\phi_{\rm P}$, can be related to the branching fractions of the semileptonic $D^+$ and $D^+_s$ decays, via
$\cot^4\phi_{\rm P}=\frac{\Gamma_{D^+_s\to\eta^\prime e^+\nu_e}/\Gamma_{D^+_s\to\eta
    e^+\nu_e}}{\Gamma_{D^+\to\eta^\prime e^+\nu_e}/\Gamma_{D^+\to\eta e^+\nu_e}}$~\cite{DiDonato:2011kr}.  
In this double ratio, both $D^+$ and $D_s^+$ differences as well as the gluonium component in the $\eta^\prime$ cancel~\cite{DiDonato:2011kr}.  
Compared with other extractions~\cite{BES:2005otq,LHCb:2014oms}, this mixing angle can give information on the gluonium component to $\eta^\prime$ state, improving our understanding of non-perturbative QCD dynamics, which is being actively explored with LQCD calculations~\cite{Christ:2010dd,Dudek:2011tt}.

Previously, the branching fractions of $D_s^+ \to \eta  e^+ \nu_e$ and $D_s^+ \to \eta^\prime  e^+ \nu_e$ were measured  by CLEO-c
\cite{Brandenburg:1995qq,Yelton:2009aa,Hietala:2015jqa} and BESIII~\cite{Ablikim:2016rqq,bes3_etaev}. Benefitting from the large data sample, BESIII reported measurements of the dynamics of these two decays, using 3.19 fb$^{-1}$ of $e^+e^-$ collision data taken at the center-of-mass energy $E_{\rm CM}=4.178$ GeV~\cite{bes3_etaev}.  
This paper reports the updated measurements of the $D^+_s\to \eta^{(\prime)} e^+\nu_e$ decay branching fractions and dynamics using 7.33~fb$^{-1}$ of $e^+e^-$ collision data collected by the BESIII detector at $E_{\rm CM}=$ 4.128 GeV, 4.157 GeV, 4.178 GeV, 4.189 GeV, 4.199 GeV, 4.209 GeV, 4.219 GeV, and 4.226 GeV. 
The integrated luminosities~\cite{BESIII:lumi} for these subsamples are 0.402~fb$^{-1}$, 0.409~fb$^{-1}$, 3.189~fb$^{-1}$, 0.570~fb$^{-1}$, 0.526~fb$^{-1}$, 0.572~fb$^{-1}$, 0.569~fb$^{-1}$, and 1.092~fb$^{-1}$, respectively, with an uncertainty of 1\%. Charge conjugated modes are implied throughout this paper.

\section{BESIII detector and Monte Carlo simulations}

The BESIII detector~\cite{Ablikim:2009aa} records symmetric $e^+e^-$ collisions
provided by the BEPCII storage ring~\cite{Yu:IPAC2016-TUYA01} in the center-of-mass energy range from 2.0 to 4.95~GeV, with a peak luminosity of $1 \times 10^{33}\;\text{cm}^{-2}\text{s}^{-1}$ 
achieved at $\sqrt{s} = 3.77\;\text{GeV}$. 
BESIII has collected large data samples in this energy region~\cite{Ablikim:2019hff,Li:2021iwf}. The cylindrical core of the BESIII detector covers 93\% of the full solid angle and consists of a helium-based
 multilayer drift chamber~(MDC), a plastic scintillator time-of-flight
system~(TOF), and a CsI(Tl) electromagnetic calorimeter~(EMC),
which are all enclosed in a superconducting solenoidal magnet
providing a 1.0~T magnetic field. The solenoid is supported by an
octagonal flux-return yoke with resistive plate counter muon
identification modules interleaved with steel.
The charged-particle momentum resolution at $1~{\rm GeV}/c$ is
$0.5\%$, and the ${\rm d}E/{\rm d}x$ resolution is $6\%$ for electrons
from Bhabha scattering. The EMC measures photon energies with a
resolution of $2.5\%$ ($5\%$) at $1$~GeV in the barrel (end cap)
region. The time resolution in the TOF barrel region is 68~ps, while
that in the end cap region was 110~ps. The end cap TOF
system was upgraded in 2015 using multi-gap resistive plate chamber
technology, providing a time resolution of
60~ps~\cite{etof}. Approximately 83\% of the data used here was collected after this upgrade.

Simulated samples produced with a {\sc
geant4}-based~\cite{geant4} Monte Carlo (MC) package, which
includes the geometric description~\cite{Huang:2022wuo} of the BESIII detector and the
detector response, are used to determine detection efficiencies
and to estimate backgrounds. The simulation models the beam
energy spread and initial state radiation (ISR) in the $e^+e^-$
annihilations with the generator {\sc
kkmc}~\cite{ref:kkmc}.
 The input cross section of $e^+e^-\to D^\pm_sD^{*\mp}_s$ is taken from Ref.~\cite{crosssection}.
The ISR production of vector charmonium(-like) states
and the continuum processes are incorporated in {\sc
kkmc}~\cite{ref:kkmc}.
In the simulation, the production of open-charm
final states directly via $e^+e^-$ annihilations is modeled with the generator {\sc conexc}~\cite{ref:conexc},
and their subsequent decays are modeled by {\sc evtgen}~\cite{ref:evtgen} with
known branching fractions from the Particle Data Group~\cite{PDG2016}.
The remaining unknown charmonium decays
are modelled with {\sc lundcharm}~\cite{ref:lundcharm}. Final state radiation
from charged final-state particles is incorporated using the {\sc
photos} package~\cite{photos}.

\section{Analysis method}

A double-tag (DT) measurement strategy, analogous to what is used in Refs.~\cite{bes3_etaev,hajime2021,DTmethod}, is employed.
At $E_{\rm CM}$ between 4.128 and 4.226~GeV, $D_s$ mesons are produced mainly from the process $e^+e^-\to D_s^{*\pm}[\to\gamma(\pi^0)D_s^\pm]D_s^\mp$. 
First, a $D_s^-$ meson is fully reconstructed in one of 
several hadronic decay modes, discussed in Sec.~\ref{sec:ST}; 
this is referred to as a single-tag (ST) candidate. 
This includes the $D_s$ directly from $e^+e^-$ annihilations and the $D_s$ from $D_s^*$ decays. Then, the signal decay of the $D_s^+$ meson and the transition $\gamma(\pi^0)$ from the $D_s^{*\pm}$ decay are reconstructed from the remaining particles in the event; these are the DT candidates. 
The branching fraction of the semileptonic decay is determined by
\begin{equation}
\mathcal B_{\rm SL}=\frac{N_{\rm DT}}{N_{\rm ST} \cdot \bar\epsilon_{\gamma(\pi^0) {\rm SL}}\cdot\mathcal B_{\rm sub}}.
\label{eq1}
\end{equation}
Here, $N_{\rm DT}=\sum\limits_{ij} N_{\rm DT}^{ij}$ and $N_{\rm ST}=\sum\limits_{ij} N_{\rm ST}^{ij}$ are the total DT and ST yields in data summing over tag mode $i$ and data set $j$;
$\bar\epsilon_{\gamma(\pi^0) {\rm SL}}$ is the efficiency of detecting the transition $\gamma(\pi^0)$ and the semileptonic decay
 in the presence of the ST $D^-_s$ candidate,
weighted by the ST yields in data.
It is calculated by $\sum\limits_{ij}\left [(N_{\rm ST}^{ij}/N_{\rm ST}) \cdot (\epsilon^{ij}_{\rm DT}/\epsilon^{ij}_{\rm ST})\right ]$, where $\epsilon^{ij}_{\rm DT}$ and $\epsilon^{ij}_{\rm ST}$ are
the detection efficiencies of the DT and ST candidates, respectively. The efficiencies do not include the branching fractions of $\eta^{(\prime)}$~\cite{PDG2022}. The quantity ${\mathcal B}_{\rm sub}$ is the product of the branching fractions of the relevant intermediate decays.

\section{Single-tag event selection}
\label{sec:ST}
The ST $D^-_s$ candidates are reconstructed from the fourteen hadronic decay modes 
$D^-_s\to 
K^+K^-\pi^-$, 
$K^- \pi^+ \pi^-$,
$\pi^+\pi^-\pi^-$,
$K^+K^-\pi^-\pi^0$, 
$\eta^\prime_{\gamma\rho^0}\pi^-$, 
$\eta_{\gamma\gamma}\rho^-$, 
$K^0_SK^-\pi^+\pi^-$,
$K^0_SK^+\pi^-\pi^-$,
$\eta_{\gamma\gamma}\pi^-$, 
$K^0_SK^0_S\pi^-$,
$\eta_{\pi^0\pi^+\pi^-}\pi^-$,
$\eta^\prime_{\eta_{\gamma\gamma}\pi^+\pi^-}\pi^-$, 
$K^0_SK^-\pi^0$,
and $K^0_SK^-$, 
where the subscripts of $\eta$ and $\eta^{\prime}$ represent the decay modes used to reconstruct these mesons. Throughout this paper, $\rho$ denotes $\rho(770)$.

The selection criteria of $K^\pm$, $\pi^\pm$, $K^0_S$, $\gamma$, $\pi^0$, and $\eta$ are the same as those used in previous works~\cite{bes2019,bes3_etaev,bes3_gev}.
All charged tracks must be within a polar angle range $|\!\cos\theta|<0.93$.
Except for those from $K^0_S$ decays, they are required to satisfy $|V_{xy}|<1$ cm and $|V_{z}|<10$~cm.
Here,
$\theta$ is the polar angle with respect to the MDC axis,
and $|V_{xy}|$ and $|V_{z}|$ are the distances of the closest approach in the transverse plane and
along the MDC axis, respectively.
The particle identification (PID) of the charged particles is performed with the combined ${\rm d}E/{\rm d}x$ and TOF information.
The combined likelihoods ($\mathcal{L}'$) under the pion and kaon hypotheses  are obtained.
Kaon and pion candidates are required to satisfy $\mathcal{L}_K>\mathcal{L}_\pi$ and $\mathcal{L}_\pi>\mathcal{L}_K$, respectively.

Each $K_{S}^0$ candidate is reconstructed from two oppositely charged tracks satisfying $|V_{z}|<$ 20~cm.
The two charged tracks are assigned
as $\pi^+\pi^-$ without imposing PID criteria. They are constrained to
originate from a common vertex and are required to have an invariant mass
within $|M_{\pi^{+}\pi^{-}} - m_{K_{S}^{0}}|<$ 12~MeV$/c^{2}$, where
$m_{K_{S}^{0}}$ is the $K^0_{S}$ nominal mass~\cite{PDG2022} and 12~MeV$/c^{2}$ corresponds to about three times the fitted resolution around the $K_S^0$ nominal mass. The
decay length of the $K^0_S$ candidate is required to be greater than
twice the vertex resolution away from the interaction point.

The $\pi^0$ and $\eta$ mesons are reconstructed from photon pairs.
Photon candidates are identified as isolated showers in the EMC. The deposited energy of each shower must be more than 25~MeV in the barrel region ($|\cos \theta|< 0.80$) and more than 50~MeV in the end cap region ($0.86 <|\cos \theta|< 0.92$). The different energy thresholds for the barrel and end cap regions are due to different energy resolutions. 
To exclude showers that originate from
charged tracks,
we require the angle subtended by the EMC shower and the position of the closest charged track at the EMC
must be greater than 10$^\circ$ as measured from the IP. 
The difference between the EMC time and the event start time, which is the interval of the trigger start time to the real collision time~\cite{starttime}, is required to be within 
(0, 700)\,ns to suppress electronic noise and showers unrelated to the event.
To form $\pi^0$ and $\eta$ candidates, we require the invariant masses of the selected photon pairs, $M_{\gamma\gamma}$, to be
within the intervals $(0.115,\,0.150)$ and $(0.500,\,0.570)$\,GeV$/c^{2}$, respectively.
To improve momentum resolution and suppress background, a kinematic fit is imposed on the selected photon pairs by constraining their invariant mass to the nominal $\pi^{0}$ or $\eta$ mass~\cite{PDG2022}.

The $\rho^0$ and $\rho^-$ candidates are reconstructed from the $\pi^+\pi^-$ and $\pi^-\pi^0$
combinations with invariant masses within the interval $(0.570,\,0.970)~\mathrm{GeV}/c^2$.

For the tag modes $D^-_s\to \eta_{\pi^0\pi^+\pi^-}\pi^-$ and $\eta_{\pi^0\pi^+\pi^-}\rho^-$,
the $\pi^0\pi^+\pi^-$ decay mode is also used to form $\eta$ candidates and
the invariant mass, $M_{\pi^0\pi^+\pi^-}$, is required to be within the interval $(0.530,\,0.570)~\mathrm{GeV}/c^2$.
To form $\eta^\prime$ candidates, two decay modes $\eta_{\gamma\gamma}\pi^+\pi^-$ and $\gamma \rho^0$ are used; 
their invariant masses are required to be within
the intervals $(0.946,\,0.970)$ GeV/$c^2$ and $(0.940,\,0.976)~\mathrm{GeV}/c^2$, respectively. The difference in the invariant mass requirements for the
$\eta_{\gamma\gamma}\pi^{+}\pi^{-}$ and $\gamma\rho^{0}$ decay modes is mainly due to different mass resolutions. 
In addition, the minimum energy
of the $\gamma$ from $\eta'\to\gamma\rho^0$ decays must be greater than 0.1\,GeV.

The momentum of any pion, which does not originate from a $K_S^0$, $\eta$, or $\eta^\prime$ decay, is required to be greater than 0.1\,GeV/$c$ to reject the soft pions from $D^{*+}$ decays. For the tag mode $D^-_s\to \pi^+\pi^-h^-$ ($h=K$ or $\pi$), the peaking background from $D^-_s\to K^0_S(\to \pi^+\pi^-)h^-$ is rejected by requiring the invariant mass of any $\pi^+\pi^-$ combination at least 30\,MeV$/c^2$ away from the nominal $K^0_S$ mass~\cite{PDG2022}.

To suppress non$\text{-}D_s^{\pm}D^{*\mp}_s$ events,
the beam-constrained mass of the ST $D_s^-$
candidate
\begin{equation}
M_{\rm BC}\equiv\sqrt{E^2_{\rm CM}/4c^4-|\vec{p}_{\rm tag}|^2/c^2}
\end{equation}
is required to be within the intervals shown in Table~\ref{tab:mbc},
where
$\vec{p}_{\rm tag}$ is the momentum of the ST $D_s^-$ candidate in the rest frame of the $e^+e^-$ initial state.
This requirement retains most of the $D_s^-$ and $D_s^+$ mesons from $e^+ e^- \to D_s^{*\mp}D_s^{\pm}$.

\begin{table}[htp]
	\centering\linespread{1.15}
	\caption{The  $M_{\rm BC}$ requirements for various energy points.}
	\small
	\label{tab:mbc}
	\begin{tabular}{cc}
		\hline\hline
		$E_{\rm CM}$ (GeV) &$M_{\rm BC}$ (GeV/$c^2$) \\
		\hline
		4.128   & $[2.010,2.061]$         \\
		4.157    &   $[2.010,2.070]$         \\
		4.178     & $[2.010,2.073]$         \\
		4.189    & $[2.010,2.076]$         \\
		4.199    & $[2.010,2.079]$         \\
		4.209    &$[2.010,2.082]$         \\
		4.219    &   $[2.010,2.085]$         \\
		4.226    &  $[2.010,2.088]$         \\
		\hline\hline
	\end{tabular}
\end{table}

\begin{figure*}[htbp]
\centering
\setlength{\abovecaptionskip}{-1pt}
\setlength{\belowcaptionskip}{-3pt}

\includegraphics[width=0.95\textwidth]
{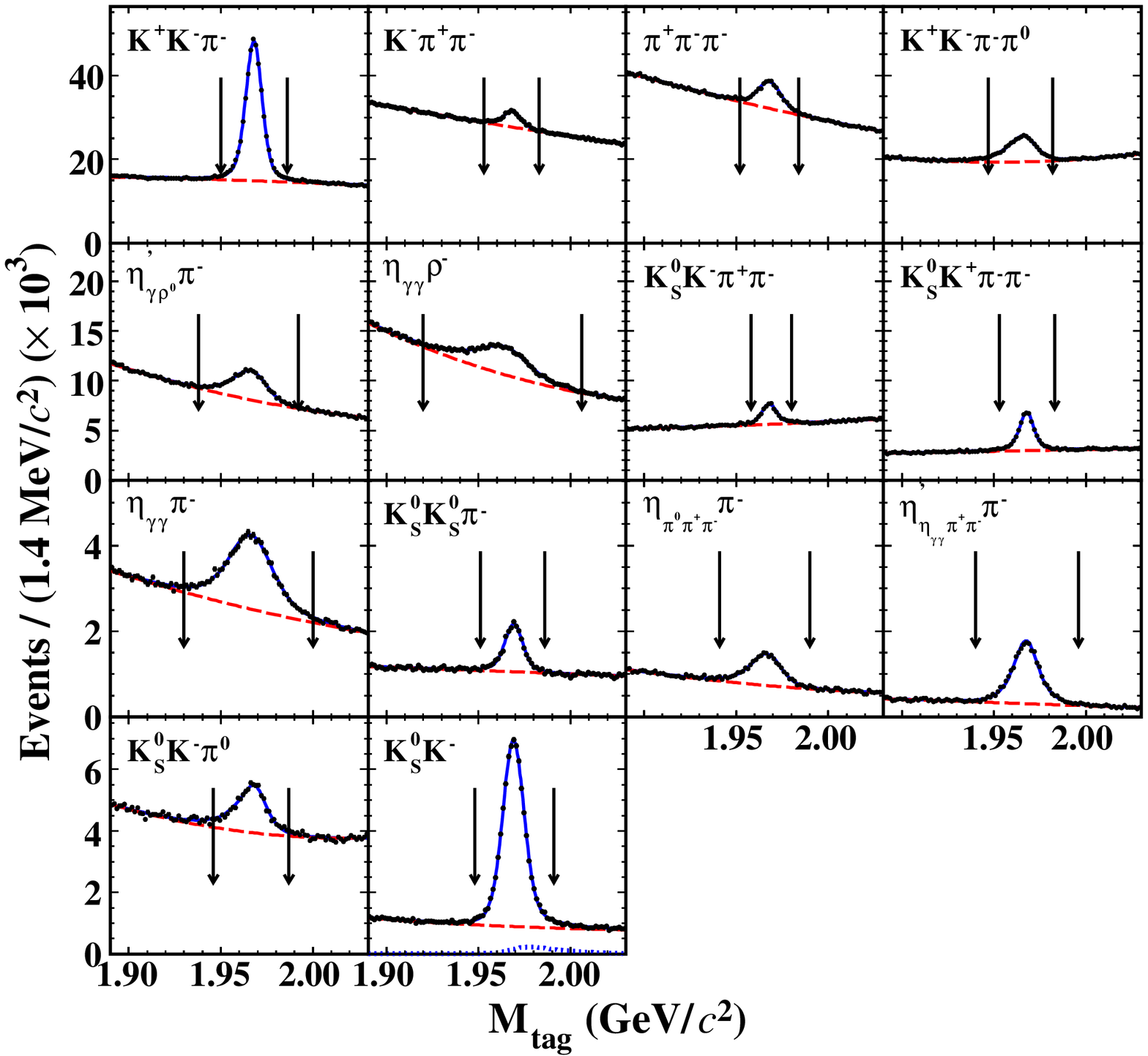}
\caption{\footnotesize
Fits to the $M_{\rm tag}$ distributions of the accepted ST candidates from the data sample with all data sets.
Points with error bars are data.
The blue solid curves are the total fit results.
The red dashed curves are the fitted backgrounds.
The blue dotted curve in the $K_S^0K^-$ mode is the $D^-\to K_S^0\pi^-$ component.
In each sub-figure, the pair of arrows denotes the signal regions.
}
\label{fig:stfit}
\end{figure*}

If there are multiple candidates for any tag mode, for a given ST $D_s$ charge, in one event,  the candidate with the $D_s^-$ recoil mass 
\begin{equation}
M_{\rm rec} \equiv \sqrt{ \left (E_{\rm CM} - \sqrt{|\vec p_{\rm tag}|^2c^2+m^2_{D^-_s}c^4} \right )^2/c^4
-|\vec p_{\rm tag}|^2/c^2}
\end{equation}
closest to the nominal $D_s^{*+}$ mass~\cite{PDG2022} is kept. Here, $m_{D_s^-}$ is the nominal  $D_s^-$ mass~\cite{PDG2022}. The probability of the best candidate selection for individual tag modes ranges in (82-99)\%.
Figure~\ref{fig:stfit} shows the invariant mass ($M_{\rm tag}$) spectra of the accepted ST candidates for
the 14 tag modes.
For each tag mode, the ST yield is obtained by a fit to
the corresponding $M_{\rm tag}$ spectrum.
The signal is described by the simulated shape for events with the angle between the reconstructed and generated four-momentum less than 15$^\circ$, convolved with a Gaussian function
representing the difference in resolution between data and simulation.
For the tag mode $D^-_s\to K_S^0K^-$,
the peaking background from $D^-\to K^0_S\pi^-$ is described by the simulated shape convolved with the same Gaussian function used in the signal shape and its yield is left as a free parameter.
The non-peaking background is modeled by a second-order Chebychev polynomial, which has been validated using the inclusive simulation sample.
The fit results for the data sample combined from all energy points are shown in Fig.~\ref{fig:stfit}.
The candidates in the signal regions, marked with black arrows in each sub-figure, are kept for further analyses. The background contributions from $e^+e^-\to(\gamma_{\rm ISR})D_s^+D_s^-$, whose contribution is (0.7-1.1)\% in the fitted ST yields for 14 tag modes based on  simulation, are subtracted in this analysis.
The resulting ST yields ($N_{\rm ST}$) for the different tag modes in data and the corresponding ST efficiencies ($\epsilon_{\rm ST}$) are summarized in the second and third columns of Table~\ref{tab:bf}, respectively. 

\begin{table*}[htbp]
\centering\linespread{1.1}
        \caption{
        The obtained values of $N_{\rm ST}$, $\epsilon_{\rm ST}$, and $\epsilon_{\rm DT}$ for various signal decays in the $i$-th tag mode, where the efficiencies do not include the branching fractions of the sub-resonant decays and the uncertainties are statistical only. The $\epsilon_{\eta^{(\prime)}}=\epsilon_{\rm DT, \eta^{(\prime)}}/\epsilon_{\rm ST}$ are the efficiencies of detecting the transition $\gamma(\pi^0)$ and signal channels in the presence of the ST $D_s^-$ candidates.}
\small
        \label{tab:bf}
        \scalebox{0.9}{
        \begin{tabular}{l c c  cc  cc cccc }\hline \hline
Tag mode  &$N_{\rm ST}$ & $\epsilon_{\rm ST}$ & $\epsilon_{\rm DT, \eta_{\gamma\gamma}}$&$\epsilon_{\eta_{\gamma\gamma}}$&$\epsilon_{\rm DT, \eta_{\pi^0\pi^+\pi^-}}$&$\epsilon_{\eta_{\pi^0\pi^+\pi^-}}$&$\epsilon_{\rm DT, \eta^\prime_{\eta\pi^+\pi^-}}$&$\epsilon_{\eta^\prime_{\eta\pi^+\pi^-}}$&$\epsilon_{\rm DT, \eta^\prime_{\gamma\rho^0}}$&$\epsilon_{\eta^\prime_{\gamma\rho^0}}$ \\

          &($\times 10^3$)&(\%)&(\%)&(\%)&(\%)&(\%)&(\%)&(\%)&(\%)&(\%)\\
$K^+K^-\pi^-$&280.7$\pm$0.9&40.87$\pm$0.03&18.47$\pm$0.05&45.19$\pm$0.12&7.11$\pm$0.04&17.39$\pm$0.10&7.94$\pm$0.03&19.43$\pm$0.08&9.45$\pm$0.05&23.13$\pm$0.11\\
$K^-\pi^+\pi^-$&35.2$\pm$1.0&45.38$\pm$0.26&20.10$\pm$0.08&44.29$\pm$0.31&7.78$\pm$0.07&17.15$\pm$0.18&8.58$\pm$0.06&18.91$\pm$0.17&10.14$\pm$0.08&22.34$\pm$0.21\\
$\pi^-\pi^+\pi^-$&72.7$\pm$1.4&51.87$\pm$0.16&22.63$\pm$0.08&43.63$\pm$0.21&8.84$\pm$0.08&17.04$\pm$0.15&9.84$\pm$0.06&18.97$\pm$0.13&11.56$\pm$0.08&22.29$\pm$0.17\\
$K^+K^-\pi^-\pi^0$&86.3$\pm$1.3&11.83$\pm$0.03&6.16$\pm$0.03&52.03$\pm$0.29&2.05$\pm$0.02&17.34$\pm$0.20&2.43$\pm$0.02&20.56$\pm$0.18&3.09$\pm$0.03&26.09$\pm$0.24\\
$\eta'_{\gamma\rho^{0}}\pi^-$&50.4$\pm$1.0&32.66$\pm$0.13&14.51$\pm$0.07&44.42$\pm$0.27&5.40$\pm$0.06&16.54$\pm$0.19&6.36$\pm$0.05&19.47$\pm$0.17&7.43$\pm$0.07&22.74$\pm$0.22\\
$\eta_{\gamma\gamma}\rho^-$&80.1$\pm$1.9&19.92$\pm$0.08&10.04$\pm$0.04&50.40$\pm$0.28&3.31$\pm$0.03&16.62$\pm$0.16&4.17$\pm$0.03&20.95$\pm$0.15&5.31$\pm$0.04&26.66$\pm$0.21\\
$K_S^0 K^-\pi^+\pi^-$&15.3$\pm$0.4&18.23$\pm$0.10&7.85$\pm$0.05&43.05$\pm$0.37&2.67$\pm$0.04&14.63$\pm$0.25&2.83$\pm$0.03&15.52$\pm$0.20&3.57$\pm$0.05&19.58$\pm$0.28\\
$K_S^0 K^+\pi^-\pi^-$&29.6$\pm$0.3&20.97$\pm$0.05&9.15$\pm$0.06&43.63$\pm$0.29&3.17$\pm$0.05&15.12$\pm$0.23&3.43$\pm$0.04&16.35$\pm$0.18&4.52$\pm$0.05&21.53$\pm$0.26\\
$\eta_{\gamma\gamma}\pi^-$&39.6$\pm$0.8&48.29$\pm$0.15&21.73$\pm$0.08&44.99$\pm$0.22&8.05$\pm$0.07&16.67$\pm$0.16&9.71$\pm$0.06&20.10$\pm$0.14&12.10$\pm$0.08&25.06$\pm$0.19\\
$K_S^0 K_S^0 \pi^-$&10.4$\pm$0.2&22.51$\pm$0.10&9.57$\pm$0.06&42.51$\pm$0.32&3.39$\pm$0.05&15.07$\pm$0.22&3.78$\pm$0.04&16.80$\pm$0.19&4.78$\pm$0.05&21.26$\pm$0.26\\
$\eta_{\pi^0\pi^+\pi^-}\pi^+$&11.7$\pm$0.3&23.32$\pm$0.11&10.38$\pm$0.06&44.53$\pm$0.34&3.79$\pm$0.05&16.27$\pm$0.23&4.36$\pm$0.04&18.68$\pm$0.20&5.60$\pm$0.06&24.03$\pm$0.27\\
$\eta'_{\eta_{\gamma\gamma}\pi^+\pi^-}\pi^-$&19.7$\pm$0.2&25.17$\pm$0.06&10.94$\pm$0.06&43.48$\pm$0.27&3.83$\pm$0.05&15.22$\pm$0.20&4.33$\pm$0.04&17.20$\pm$0.17&5.64$\pm$0.06&22.39$\pm$0.24\\
$K_S^0 K^-\pi^0$&23.0$\pm$0.6&16.98$\pm$0.09&8.15$\pm$0.05&47.99$\pm$0.41&2.75$\pm$0.04&16.18$\pm$0.27&3.48$\pm$0.04&20.52$\pm$0.25&4.25$\pm$0.05&25.04$\pm$0.33\\
$K_S^0 K^-$&62.2$\pm$0.4&47.36$\pm$0.06&20.46$\pm$0.08&43.19$\pm$0.18&7.95$\pm$0.07&16.78$\pm$0.15&9.07$\pm$0.06&19.15$\pm$0.13&10.69$\pm$0.08&22.56$\pm$0.17\\

Average&&&&45.93$\pm$0.07&&16.86$\pm$0.05&&19.39$\pm$0.04&&23.59$\pm$0.06\\

\hline \hline
        \end{tabular}
        }
\end{table*}

\section{Double-tag event selection}

The transition photon or $\pi^0$ and the semileptonic $D^+_s$ decay candidate 
are selected from the particles remaining after ST reconstruction.  
The photon or $\pi^0$ providing the lowest energy difference, $\Delta E$, is selected.  Here,  $\Delta E = E_{\rm CM} - E_{\rm tag} - E_{\gamma(\pi^0) \, D^-_s}^{\rm rec} - E_{\gamma(\pi^0)}$, where the recoil energy is calculated from the momenta of $\gamma(\pi^0)$ and $D_s^-$ tag as 
$E_{\gamma(\pi^0) \, D^-_s}^{\rm rec} = \sqrt{|-\vec{p}_{\gamma(\pi^0)}-\vec{p}_{\rm tag}|^2c^2 +  m^2_{D_s^+}c^4}$.  
The signal candidates are examined using the kinematic variable 
$ M_{\rm miss}^2 \equiv  \left (E_{\rm CM} - \sum\limits_i E_i\right )^2/c^4 - \left |\sum\limits_i \vec{p}_{i}\right |^2/c^2,$
 where
 $E_i$ and $\vec{p}_i$, with $i = ($tag, $\gamma(\pi^0)$, $e$ or $\eta^{(\prime)}$), are the energies and momenta of particle $i$.
To improve the $M_{\rm miss}^2$ resolution, all the selected candidate tracks in the tag side, transition $\gamma(\pi^0)$, and $\eta^{(\prime)}e^+$ of the signal side, plus the missing neutrino, are subjected to a kinematic fit with a net three constraints: seven are applied and the neutrino four-vector is determined.  The fit requires energy and momentum conservation, and in addition, the invariant masses of the two $D_s$ mesons are constrained to the nominal $D_s$ mass, the invariant mass of the $D_s^-\gamma(\pi^0)$ or $D_s^+\gamma(\pi^0)$ combination is constrained to the nominal $D_s^*$ mass, and the combination with the smaller $\chi^2$ is kept. To suppress the background contributions from non-$D_s D^*_s$ events in $D_s^+\to\eta^\prime_{\gamma\rho^0}e^+\nu_e$, the $\chi^2$ is required to satisfy $\chi^2<200$.

In the signal side, the $\eta$ meson is reconstructed by $\eta\to\gamma\gamma$ or $\pi^0\pi^+\pi^-$,  and the $\eta^\prime$ meson is reconstructed by $\eta^\prime\to\eta_{\gamma\gamma}\pi^+\pi^-$ or $\gamma\rho^0_{\pi^+\pi^-}$.
The selection criteria of  $\eta^{(\prime)}$ are the same as in the ST selection.
The positron candidate is identified by using the ${\rm d}E/{\rm d}x$, TOF, and EMC information.
Combined likelihoods for the pion, kaon and positron hypotheses, $\mathcal{L}'_\pi$, $\mathcal{L}'_K$ and $\mathcal{L}'_e$, are calculated.
Charged tracks satisfying $\mathcal{L}'_e>0.001$ and $\mathcal{L}'_e/(\mathcal{L}'_e+\mathcal{L}'_\pi+\mathcal{L}'_K)>0.8$ are assigned as positron candidates.
To suppress background contributions from $D^+_s$ hadronic decays, the maximum energy of the
unused showers ($E_{\rm \gamma~extra }^{\rm max}$) must be less than 0.3~GeV and events with
additional charged tracks ($N^{\rm extra}_{\rm char}$) are removed. The invariant mass of the $\eta^{(\prime)}$ and $e^+$ is required to be
$M_{\eta^{(\prime)} e^+}< 1.9$~GeV/$c^2$ for $D_s^+\to\eta^{(\prime)} e^+\nu_e$ to further
suppress the background contributions of $D_s^+\to\eta^{(\prime)} \pi^+$.
To suppress contributions of backgrounds to $D_s^+\to\eta^\prime_{\gamma\rho^0}e^+\nu_e$ where the photon is from a $\pi^0$ decay, the opening angle between the missing momentum and the most energetic unused shower ($\theta_{\gamma,~\rm miss}$) is required to satisfy $\cos\theta_{\gamma,~\rm miss}<0.85$.

\section{Branching fractions}

\subsection{Results of branching fractions}

\begin{figure*}[htbp]
\centering
\setlength{\abovecaptionskip}{-7pt}
\setlength{\belowcaptionskip}{-3pt}
\includegraphics[width=0.9\textwidth]{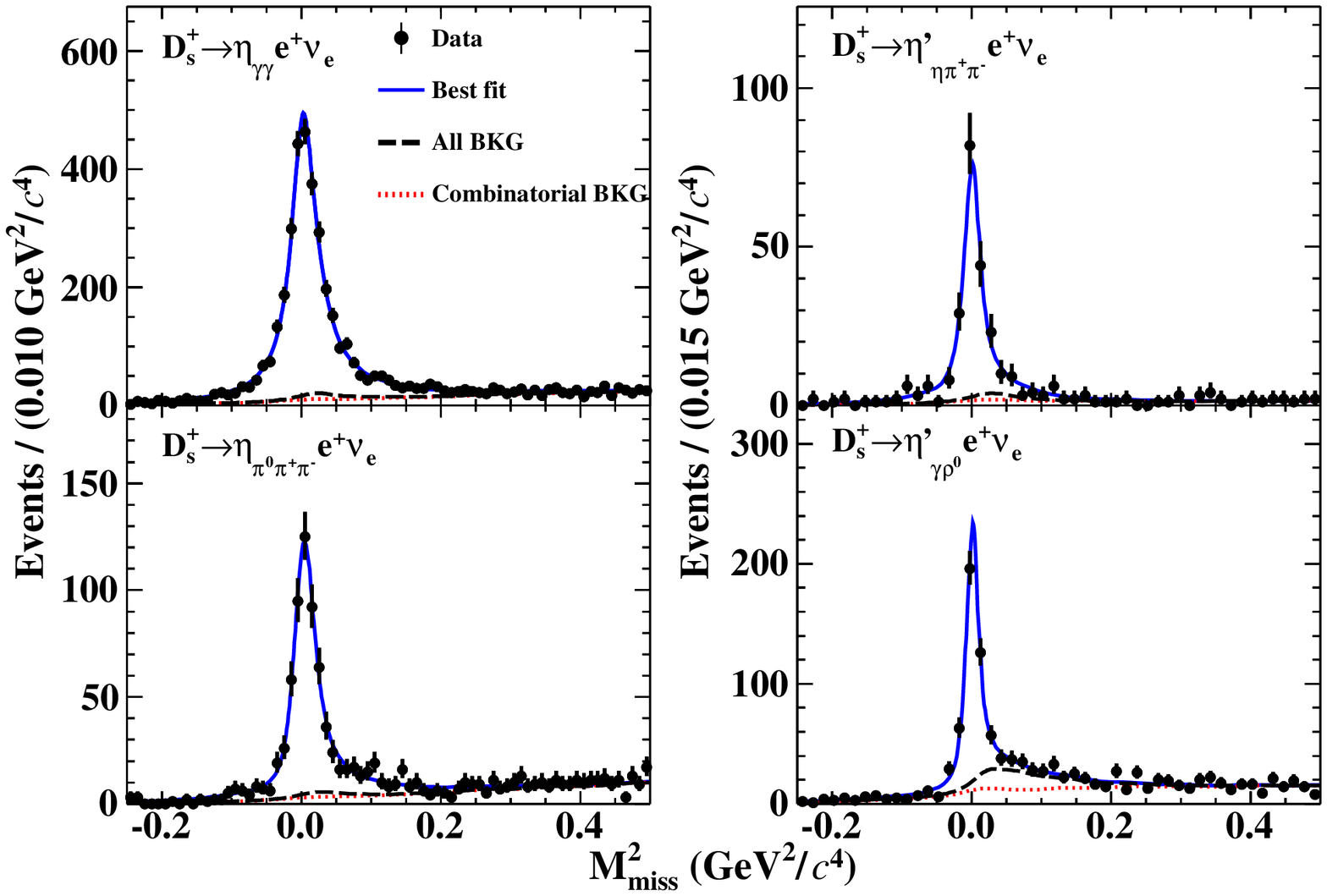}
\caption{Fits to the $M_{\rm miss}^2$ distributions of the candidate events for various semileptonic decays.
The points with error bars represent data. The blue solid curves denote the total fits, and the red solid dotted curves show
the fitted combinatorial background contributions. Differences between dashed and dotted curves are the backgrounds from $D_s^+\to\eta^{(\prime)}\pi^+\pi^0$, $\eta^{(\prime)}\mu^+\nu_\mu$, and $D_s^+ \to \phi(1020)_{\pi^0\pi^+\pi^-} e^+ \nu_e$.
}
\label{fig:fit_ep}
\end{figure*}

After imposing all selection criteria, the $M_{\rm miss}^2$ distributions of the accepted candidates for the semileptonic decays $D_s^+ \to \eta^{(\prime)} e^+ \nu_e$ are obtained,
as shown in Fig.~\ref{fig:fit_ep}.
For the semileptonic $D^+_s$ decays reconstructed via two different $\eta^{(\prime)}$ decays,
 a simultaneous unbinned maximum-likelihood fit is performed on the two $M_{\rm miss}^2$ distributions, 
  where the branching fractions of $D_s^+\to\eta^{(\prime)} e^+\nu_e$ measured with the different $\eta^{(\prime)}$ decay modes are constrained to be equal. 
The signal and background components are modeled with shapes derived from MC simulation.
The yields of the peaking backgrounds due to $D_s^+\to\eta^{(\prime)}\pi^+\pi^0$ and $\eta^{(\prime)}\mu^+\nu_\mu$ are fixed according to the MC simulation.
For $D_s^+ \to \eta^\prime_{\gamma\rho^0} e^+ \nu_e$, there is a remaining background contribution from $D_s^+ \to \phi(1020)_{\pi^0\pi^+\pi^-} e^+ \nu_e$. The yield of $D_s^+ \to \phi(1020)_{\pi^0\pi^+\pi^-} e^+ \nu_e$ is left free in the fit.
The remaining combinatorial backgrounds are dominated by open charm (more 
than 60\%) and $e^+e^- \to q \bar{q}$ (about 30\%). 
The magnitude of this contribution is a free parameter in the fit.
The branching fractions of the intermediate decays are
 ${\mathcal B}(\eta \to \gamma\gamma)=(39.36\pm0.18)\%$, ${\mathcal B}(\eta \to \pi^0\pi^+\pi^-)=(23.02\pm0.25)\%$, ${\mathcal B}(\eta^\prime\to \eta\pi^+\pi^-)=(42.5\pm0.5)\%$, ${\mathcal B}(\eta^\prime\to \pi^+\pi^-\gamma)=(29.5\pm0.4)\%$, and ${\mathcal B}(\pi^0 \to \gamma\gamma)=(98.823\pm0.034)\%$~\cite{PDG2022}.
The branching fractions of $D_s^+\to\eta^{(\prime)} e^+\nu_e$, the yields of other background contributions and the 
parameters of the Gaussian functions convolved with the distributions from MC simulation are left free during the fit.
The branching fractions are calculated from the signal yields with Eq.~\ref{eq1}. The signal efficiencies, the signal yields, and the obtained branching fractions are summarized in Table~\ref{tab:BFslep}. 

\begin{table}[hbtp]
\centering
\caption{\small Signal efficiencies ($\epsilon_{\rm \gamma(\pi^0){\rm SL}}$), signal yields ($N_{\rm DT}$), and obtained branching fractions ($\mathcal{B}_{\rm SL}$) for various semi-electronic decays.  Efficiencies include the branching fractions of $D_s^{*\mp}$ decays but do not include the branching fractions of the $\eta^{(\prime)}$ decays.
Numbers in the first and second parentheses are the statistical and systematic uncertainties, respectively.  \label{tab:BFslep}}
              \begin{tabular}{lcccc}
\hline\hline
Decay       &$\eta^{(\prime)}$ decays& $\epsilon_{\gamma(\pi^0)\rm SL}$ (\%) & $N_{\rm DT}$                      & $\mathcal B_{\rm SL}$ (\%)  \\ \hline
\multirow{2}{*}{$\eta e^+\nu_e$ }&    $\gamma\gamma$            & $45.93(07)$         & \multirow{2}{*}{4036(71) }&\multirow{2}{*}{2.255(39)(51)}\\
       &  $\pi^0\pi^+\pi^-$      & 16.86(05)         &                                &                                    \\
       \hline
\multirow{2}{*}{$\eta^\prime e^+\nu_e$} &$\eta\pi^+\pi^-$  & 19.39(04)      &\multirow{2}{*}{675(32) }&\multirow{2}{*}{0.810(38)(24)} \\
    &$\gamma\rho^0$ & 23.59(06)         &                                &                                    \\
\hline\hline
\end{tabular}
        
\end{table}

\subsection{Systematic uncertainties}

Table~\ref{sys} summarizes the sources of the systematic uncertainties in the measurements of the branching fractions of
$D_s^+\to \eta^{(\prime)}e^+\nu_e$. They are assigned relative to the measured branching fractions and are discussed below. In this table, the contributions to the systematic uncertainties listed in the upper part are treated as correlated, while those in the lower part are treated as uncorrelated.   

 The total systematic uncertainties of the branching fractions of $D_s^+\to \eta e^+\nu_e$ and $D_s^+\to \eta^\prime e^+\nu_e$ are calculated  to be 2.3\% and 2.9\%, respectively, after taking into account correlated and uncorrelated systematic uncertainties and using the method described in Ref.~\cite{Schmelling:1994pz}.

\begin{table*}[htp]
\centering
\caption{Relative systematic uncertainties (in \%) on the measurements of the branching fractions
of $D_s^+\to \eta e^+\nu_e$ and $D_s^+\to \eta^\prime e^+\nu_e$. The top and the bottom sections are correlated and uncorrelated, respectively. The uncertainty in the uncorrelated $\pi^\pm$ tracking is obtained as the square root of the quadratic difference of the total uncertainty in the $\pi^\pm$ tracking and the correlated portion. 
}
\begin{tabular}{lcc|cc}
  \hline
  \hline
  Source  & $\eta_{\gamma\gamma}e^+\nu_e$&$\eta_{\pi^0\pi^+\pi^-}e^+\nu_e$&$\eta^\prime_{\eta\pi^+\pi^-}e^+\nu_e$&$\eta^\prime_{\gamma\rho^0}e^+\nu_e$  \\
  \hline
ST $D^{-}_s$ yields                                      &0.5  &0.5  &0.5 &0.5    \\
  $\pi^0$ and $\eta$ reconstruction                                  &1.1  &1.1  &0.8 &--    \\
   $\pi^\pm$ tracking                                           &--&1.2&0.6&0.6\\
  $\pi^\pm$ PID                                                &--&0.4&0.4&0.4\\
  $e^+$ tracking                                             &0.5 &0.5 & 0.5&0.5        \\
  $e^+$ PID                                                  &0.2 &0.2 & 0.2&0.2        \\
 Transition $\gamma(\pi^0)$ reconstruction      &1.0&1.0&1.0&1.0\\
 Smallest $|\Delta E|$                                            &1.0&1.0&1.0&1.0\\
Peaking background &0.4&0.4&1.0&1.0\\
  Hadronic transition form factors                                         &0.5&0.5&0.2 &0.2        \\
\hline
   $\pi^\pm$ tracking                                           &--&--&1.7&--\\
$\eta^{(\prime)}$ selection &--&0.1&0.1&1.4\\
  Tag bias                                                     &0.4 &0.1 &0.1&0.1         \\
  $\chi^2$ requirement&--&--&--&1.5\\
  $M_{\rm \eta^{(\prime)}e^+}$ requirement                                 &neglected&neglected&neglected&neglected         \\
    $\cos\theta_{\gamma,~\rm miss}$ requirement                              &neglected&neglected&neglected&neglected              \\

  $E^{\rm max}_{\rm extra \gamma}$ and $N_{\rm extra}^{\rm char}$ requirements &0.7&2.0&2.0&1.1      \\
  $M_{\rm miss}^2$ fit                                                    & 0.4&0.7&0.7&0.5          \\
  MC  statistics                                               &0.1     &0.3&0.2&0.2           \\
  Quoted branching fractions                                   &0.5&1.1&1.3&1.4\\
  \hline
  Total                                                        &2.3           &3.4  &3.7&    3.4    \\

  \hline
  \hline
\end{tabular}
\label{sys}
\end{table*}

\paragraph{ST $D^{-}_s$ yields}

The uncertainty of the fits to the $D_s^-$ invariant mass spectra is estimated by varying the signal and background shapes and repeating the fits for both data and MC sample.
A variation of the signal shape is obtained by modifying the matching requirement between generated and reconstructed angles from 15$^\circ$ to 10$^\circ$ or 20$^\circ$.
The  background shape is changed to a third-order Chebychev polynomial.
The relative change of the ST yields in data over the ST efficiencies is considered as the systematic uncertainty.
Moreover, an additional uncertainty due to the background fluctuation of the fitted ST yields is included. 
The quadrature sum of these three terms, 0.5\%, is assigned as the associated systematic uncertainty.

\paragraph{$\pi^0$ and $\eta$ reconstruction}
\label{pi0}

The systematic uncertainty in the $\pi^0$ reconstruction has been studied
by using the control sample of $e^+e^-\to K^+K^-\pi^+\pi^-\pi^0$.
The systematic uncertainty in the $\eta$ reconstruction is taken as equal to that of the $\pi^{0}$ due to the limited $\eta$ sample.
After correcting differences of the $\pi^0$ or $\eta$ reconstruction efficiencies  between data and MC simulation,
which are $0.991-1.024$, the systematic uncertainties, due to statistical uncertainties on these corrections, are listed in Table~\ref{sys}. 

\paragraph{$\pi^\pm$ tracking and PID efficiencies}

The tracking and PID efficiencies of $\pi^\pm$ are studied by using the control sample of $e^+e^-\to K^+K^-\pi^+\pi^-$. The momentum weighted data-MC differences due to $\pi^\pm$ tracking efficiencies range from 
$0.981-1.001$ for different signal decays and
 the signal efficiencies are corrected by these factors.
The systematic uncertainties due to $\pi^\pm$ tracking and $\pi^\pm$ PID are listed in Table~\ref{sys}. The uncertainties of $\pi^\pm$ tracking for $D_s^+\to \eta^\prime_{\eta\pi^+\pi^-} e^+\nu_e$ and $D_s^+\to \eta^\prime_{\gamma\rho^0} e^+\nu_e$ are partly correlated, the common uncertainty of 0.6\% is considered as fully correlated, and the remaining quadratic difference of 1.7\% for $D_s^+\to \eta^\prime_{\eta\pi^+\pi^-} e^+\nu_e$ is uncorrelated.

\paragraph{$e^+$ tracking and PID efficiencies}
\label{sec:electron}

The $e^+$ tracking and PID efficiencies are studied by using the control sample of $e^+e^-\to \gamma e^+e^-$.
 The ratios are 
$1.000\pm0.005$ for  $e^+$ tracking and $0.988\pm0.002$ for $e^+$ PID efficiencies.  After corrections, the systematic uncertainties, due to statistical uncertainties on these factors, are listed in Table~\ref{sys}.

\paragraph{Transition $\gamma(\pi^0)$ reconstruction}
\label{sec:gamma}

The systematic uncertainty of the transition $\gamma(\pi^0)$ selection is assigned as 1.0\% based on studies of the control sample of $J/\psi\to \pi^0\pi^+\pi^-$~\cite{Ablikim:2011kv}.

\paragraph{Smallest $|\Delta E|$}

The systematic uncertainty of selecting the transition $\gamma(\pi^0)$ with the smallest $|\Delta E|$ method
is studied by using two control samples of $D_s^+\to K^+K^-\pi^+$ and $D_s^+\to \eta\pi^0\pi^+$.
The difference of the efficiency of selecting the transition $\gamma(\pi^0)$ candidates in data versus the simulation is 1.0\%, which is assigned as the systematic uncertainty.

\paragraph{Peaking background}
The systematic uncertainty due to the peaking backgrounds from $D_s^+\to\eta^{(\prime)}\pi^+\pi^0$ and $D_s^+\to\eta^{(\prime)}\mu^+\nu_\mu$ is estimated by varying the quoted branching fractions~\cite{PDG2022} by $\pm 1\sigma$ and correcting by the data-MC difference for the mis-identification of $\pi^+\to e^+$ and $\mu^+\to e^+$.  
The relative changes of signal yields are taken as  the corresponding systematic uncertainties and listed in Table~\ref{sys}.

\paragraph{Hadronic transition form factors}

The detection efficiencies are estimated by using signal MC events generated with the hadronic transition form factors measured in this work.
The corresponding systematic uncertainties are estimated by varying the parameters by $\pm1\sigma$ and listed in Table~\ref{sys}.

\paragraph{$\eta^{(\prime)}$ selection}
The systematic uncertainties due to the $\eta^{(\prime)}$ invariant mass requirements are estimated to be 0.1\%, 0.1\%, and 1.0\% for $D_s^+\to\eta_{\pi^0\pi^+\pi^-} e^+\nu_e$, $D_s^+\to\eta^\prime_{\eta\pi^+\pi^-} e^+\nu_e$, and $D_s^+\to\eta^\prime_{\gamma\rho^0} e^+\nu_e$, respectively, by analyzing the difference of the resolution of $M_{\eta^{(\prime)}}$ between data and MC simulation with the sample $J/\psi\to\phi\eta^{(\prime)}$. 
Additionally, a 1.0\% uncertainty, related to the $\gamma$ reconstruction efficiency in the $\eta^\prime \to  \gamma \rho^0$ decay, is estimated by studying a control sample 
 of $J/\psi\to \pi^0\pi^+\pi^-$~\cite{Ablikim:2011kv}.

\paragraph{Tag bias}

Due to different reconstruction environments in the inclusive and signal MC samples,
the ST efficiencies determined by the inclusive MC sample may be different from those by the signal MC sample.
This may lead to incomplete cancellation of the systematic uncertainties associated with the ST selection, referred to as ``tag bias''.  
Inclusive and signal MC efficiencies are compared and the tracking and PID efficiencies for kaons and pions are studied for different track multiplicities.  The resulting ST-average offsets are assigned as the systematic uncertainties from tag bias and listed in Table~\ref{sys}.

\paragraph{$\chi^2$ requirement}

The systematic uncertainty due to the $\chi^2$ requirement is estimated with a hadronic DT sample with $D^+_s\to \eta^\prime_{\gamma\rho^0}\pi^+$ replacing the semileptonic signal. The difference of the accepted efficiencies of the $\chi^2$ requirement between data and MC simulation is 1.5\%, which is assigned as the systematic uncertainty for $D_s^+\to \eta^\prime_{\gamma\rho^0} e^+\nu_e$.

\paragraph{$M_{\eta^{(\prime)}e^+}$ requirement}
\label{m}

The efficiencies of the $M_{\eta^{(\prime)}e^+}<1.9$~GeV$/c^2$ requirement are greater than 99\% for all signal decays and the differences of these efficiencies between data and MC simulation are negligible.

\paragraph{$\cos\theta_{\gamma,~\rm miss}$ requirement}

The systematic uncertainty due to the $\cos\theta_{\gamma,~\rm miss}$ requirement is estimated by varying the requirement by $\pm0.05$.
The differences of the branching fractions are negligible.

\paragraph{$E_{\rm extra\gamma}^{\rm max}$ and $N_{\rm extra}^{\rm char}$ requirements}
\label{em}

The systematic uncertainty in the $E_{\rm extra \gamma}^{\rm max}$ and $N_{\rm extra}^{\rm char}$ requirements is estimated with a hadronic DT sample with $D^+_s\to \eta_{\gamma\gamma}\pi^+$, $\eta_{\pi^0\pi^+\pi^-}\pi^+$, $\eta^\prime_{\eta\pi^+\pi^-}\pi^+$, and $\eta^\prime_{\gamma\rho^0}\pi^+$.
The associated systematic uncertainties are listed in Table~\ref{sys}.

\paragraph{$M_{\rm miss}^2$ fit}

The systematic uncertainty due to the $M_{\rm miss}^2$ fit is considered in two parts.
Since a Gaussian function is convolved with the simulated signal shapes to account for the resolution difference between
data and MC simulation, the systematic uncertainty from the signal shape is ignored.
The systematic uncertainty due to the background shape is assigned by varying
the relative fractions of major backgrounds from $e^+e^-\to q\bar q$ and non-$D_s^{*\pm}D_s^\mp$ open-charm processes within $\pm$30\% according to the uncertainty of its input cross section in the inclusive MC sample.
The changes in the branching fractions are taken as  the corresponding systematic uncertainties and listed in Table~\ref{sys}.

\paragraph{MC statistics}

The relative uncertainties of the signal efficiencies are assigned as the systematic uncertainties due to MC statistics, as listed in Table~\ref{sys}.

\paragraph{Quoted branching fractions}

The uncertainties in the quoted branching fractions of $\pi^0\to\gamma\gamma$, $\eta\to\gamma\gamma$, $\eta\to\pi^0\pi^+\pi^-$ $\eta^\prime\to\eta\pi^+\pi^-$, and $\eta^\prime\to \pi^+\pi^-\gamma$ are 0.03\%, 0.5\%, 1.1\%,  1.2\%, and 1.4\%, respectively.
The quoted branching fraction of $D_s^{*+}\to \pi^0 D_s^+$ is measured relative to $D_s^{*+}\to\gamma D_s^+$. Thus, they are fully correlated with each other
and their uncertainty is 0.7\%. 
The change in signal detection efficiency when changing these branching fractions by $\pm 1 \sigma$ is at most 0.1\%, which is assigned as the systematic uncertainty.  Quadratically summing these two effects gives the associated systematic uncertainties 0.5\%, 1.1\%, 1.3\%, and 1.4\% for
$D_s^+\to\eta_{\gamma\gamma} e^+\nu_e$, $D_s^+\to\eta_{\pi^0\pi^+\pi^-} e^+\nu_e$, $D_s^+\to \eta^\prime_{\eta\pi^+\pi^-} e^+\nu_e$, and $D_s^+\to \eta^\prime_{\gamma\rho^0} e^+\nu_e$, respectively.

\section{Hadronic transition form factors}

The differential decay width can be expressed as
\begin{equation}
    \frac{\mathrm{d}\Gamma\left(D_s^+\to\eta^{(\prime)}e^+\nu_e\right)}{\mathrm{d}q^2} = \frac{G_F^2|V_{cs}|^2}{24\pi^3}\left |f_+^{\eta^{(\prime)}}(q^2)\right |^2\left |\vec p_{\eta^{(\prime)}}\right |^3,
\end{equation}
where $q$ is the momentum transfer to the $e^+\nu_e$ system, $|\vec p_{\eta^{(\prime)}}|$ is the magnitude of the meson 3-momentum in the $D_s^+$ rest frame and $G_F$ is the Fermi constant.
In the modified pole model~\cite{Becher:2005bg},
\begin{equation}
    f^{\eta^{(\prime)}}_+(q^2) = \frac{f^{\eta^{(\prime)}}_+(0)}{\left(1-\frac{q^2}{M^2_{\rm pole}}\right)\left(1-\alpha\frac{q^2}{M^2_{\rm pole}}\right)},
    \label{eq:mod_pole}
\end{equation}
where $M_{\rm pole}$ is fixed to $m_{D^{*+}_s}$ and $\alpha$ is a free parameter. The simple pole model~\cite{Becirevic:1999kt} is obtained by setting $\alpha=0$ and leaving $M_{\rm pole}$ free.
In the two-parameter (2-Par) series expansion~\cite{Becher:2005bg},
the hadronic transition form factor is given by
\begin{equation}
\begin{array}{l}
	f^{\eta^{(\prime)}}_{+}(q^2)=\frac{1}{P(q^2)\Phi(q^2)}\frac{f^{\eta^{(\prime)}}_{+}(0)P(0)\Phi(0)}{1+r_{1}(t_{0})z(0,t_{0})}\\
	\displaystyle\times\left(1+r_{1}(t_{0})[z(q^2,t_{0})]\right).
	\end{array}
\end{equation}
Here, $P(q^2)=z(q^2,m_{D^{*}_{s}}^{2})$, where $z(q^2,t_{0})=\frac{\sqrt{t_{+}-q^2}-\sqrt{t_{+}-t_{0}}}{\sqrt{t_{+}-q^2}+\sqrt{t_{+}-t_{0}}}$. $\Phi$ is given by
\begin{equation}
\begin{array}{l}
	\displaystyle \Phi(q^2)=\sqrt{\frac{1}{24\pi\chi_{V}}}\left(\frac{t_{+}-q^2}{t_{+}-t_{0}}\right)^{1/4}\left(\sqrt{t_{+}-q^2}+\sqrt{t_{+}}\right)^{-5}\\
	\displaystyle \times\left(\sqrt{t_{+}-q^2}+\sqrt{t_{+}-t_{0}}\right)\left(\sqrt{t_{+}-q^2}+\sqrt{t_{+}-t_{-}}\right)^{3/2}\\
	\displaystyle \times\left(t_{+}-q^2\right)^{3/4},
\end{array}
\end{equation}
where $t_{\pm}=(m_{D^+_s}\pm m_{\eta^{(\prime)}})^{2}$, $t_{0}=t_{+}(1-\sqrt{1-t_{-}/t_{+}})$,
$m_{D^+_s}$ and $m_{\eta^{(\prime)}}$ are the masses of $D^+_s$ and $\eta^{(\prime)}$ particles,
$m_{D_s^*}$ is the pole mass of the vector form factor accounting for the strong interaction between
$D^+_s$ and $\eta^{(\prime)}$ mesons and usually taken as the mass
of the lowest lying $c\bar s$ vector meson $D_s^*$~\cite{PDG2022},
 and $\chi_{V}$ is obtained from dispersion
relations using perturbative QCD~\cite{chiV}.

\subsection{Differential decay rates}

To extract the hadronic transition form factors of the semileptonic decays, the differential decay rates are measured in different $q^{2}$ intervals.
For the $D_s^+ \to \eta e^+ \nu_e$ decay, the $q^2$ range $(m_e^2, 2.02)$ GeV$^2/c^4$ is subdivided in eight intervals of 0.2 GeV$^2/c^4$ width (except for a wider final bin), while three regions, $(m_e^2, 0.3)$, (0.3,0.6), and (0.6,1.02) GeV$^2/c^4$, are defined for $D_s^+ \to \eta^\prime e^+ \nu_e$.
The differential decay rates in the individual $q^2$ intervals $i$ are determined as
\begin{equation}
	\frac{\mathrm{d}\Gamma_{i}}{\mathrm{d}q_{i}^{2}}=\frac{\Delta\Gamma_{i}}{\Delta q^{2}_{i}},
\end{equation}
where 
	$\Delta\Gamma_{i}=\frac{N_{\mathrm{prd}}^{i}}{\tau_{D_s^+}\cdot N_{\mathrm{ST}}}$
	is the decay rate in the $i$-th $q^2$ interval, $N_{\mathrm{prd}}^{i}$ is the number of events produced in the $i$-th $q^{2}$ interval,
$\tau_{D^+_s}$ is the $D_s^+$ lifetime~\cite{PDG2022} and $N_{\mathrm{ST}}$ is the number of the ST $D_s^-$ mesons.

In the $i$-th $q^{2}$ interval, the number of events produced in data is calculated as
\begin{equation}
	N_{\mathrm{prd}}^{i}=\sum_{j}^{N_{\mathrm{intervals}}}\left(\varepsilon^{-1}\right)_{ij}N_{\mathrm{DT}}^{j},
\end{equation}
	where $(\varepsilon^{-1})_{ij}$ is the element of the inverse efficiency matrix, obtained by analyzing the signal MC events.
The statistical uncertainty of $N_{\mathrm{prd}}^{i}$ is given by
\begin{equation}
\left[\sigma\left(N_{\mathrm{prd}}^{i}\right)\right]^2=\sum_{j}^{N_{\mathrm{intervals}}}\left(\varepsilon^{-1}\right)_{ij}^2\left[\sigma_{\rm stat}\left(N_{\mathrm{DT}}^{j}\right)\right]^2,
\end{equation}
where $\sigma_{\rm stat}(N_{\mathrm{DT}}^{j})$ is the statistical uncertainty of $N_{\mathrm{DT}}^{j}$.
The efficiency matrix $\varepsilon_{ij}$ is given by
\begin{equation}
	\varepsilon_{ij}=\frac{N_{ij}^{\mathrm{rec}}}{N_{j}^{\mathrm{gen}}}\cdot \frac{1}{\varepsilon_{\mathrm{tag}}}\cdot f_{j}^{\mathrm{corr}},
\end{equation}
	where $N_{ij}^{\mathrm{rec}}$ is the number of events generated in the $j\text{-}$th $q^{2}$ interval and reconstructed in the $i$-th $q^{2}$ interval, $N_{j}^{\mathrm{gen}}$ is the total number of events generated in the $j\text{-}$th $q^{2}$ interval, and $\varepsilon_{\mathrm{tag}}$ is the ST efficiency. $f_{j}^{\mathrm{corr}}$ is the efficiency correction factor for the events generated in the $j$-th $q^{2}$ interval, which is obtained with the same analysis procedure as that in the branching fraction measurement.
The product of the efficiency correction factors in each $q^2$ is listed in Table~\ref{fcorr}.

\begin{table*}
\centering
\caption{Summary of efficiency correction factors, $f_{\rm corr}^i$, in each $q^2$ bin.}
\begin{tabular}{ccccc}
\hline\hline
$q^2$&$D_s^+\to\eta_{\gamma\gamma}e^+\nu_e$&$D_s^+\to\eta_{\pi^0\pi^+\pi^-}e^+\nu_e$&$D_s^+\to\eta^\prime_{\eta\pi^+\pi^-}e^+\nu_e$&$D_s^+\to\eta^\prime_{\gamma\rho^0}e^+\nu_e$\\
\hline
1&0.980$\pm$0.018&0.996$\pm$0.012&0.967$\pm$0.014&0.981$\pm$0.005\\
2&0.978$\pm$0.016&0.997$\pm$0.012&0.971$\pm$0.015&0.980$\pm$0.005\\
3&0.976$\pm$0.013&0.996$\pm$0.013&0.976$\pm$0.017&0.979$\pm$0.005\\
4&0.973$\pm$0.011&0.997$\pm$0.014&-&-\\
5&0.971$\pm$0.010&0.998$\pm$0.016&-&-\\
6&0.974$\pm$0.009&0.998$\pm$0.017&-&-\\
7&0.978$\pm$0.009&0.998$\pm$0.019&-&-\\
8&0.990$\pm$0.009&0.996$\pm$0.023&-&-\\
\hline\hline
\end{tabular}
\label{fcorr}
\end{table*}

Tables~\ref{tab:effmatrixa} and \ref{tab:effmatrixb} give the elements of the efficiency matrices weighted by the ST yields in the data sample.

\begin{table*}[htbp]\centering
\caption{The efficiency matrices for $D_s^+\to\eta e^+\nu_e$ averaged over all 14 ST modes, where $\varepsilon_{ij}$ represents the
efficiency in \% for events produced in the $j$-th $q^2$ interval and reconstructed in the $i$-th $q^2$ interval. Efficiencies do not include the branching fractions of $\eta^{(\prime)}$ decays.}
\label{tab:effmatrixa}
\begin{tabular}{c|cccccccc|cccccccc}\hline\hline
\multirow{2}{*}{$\varepsilon_{ij}$}&\multicolumn{8}{c|}{$D_s^+\to\eta_{\gamma\gamma}e^+\nu_e$}&\multicolumn{8}{c}{$D_s^+\to\eta_{\pi^0\pi^+\pi^-}e^+\nu_e$}\\
&1&2&3&4&5&6&7&8&1&2&3&4&5&6&7&8\\
\hline
1&43.65&4.46&0.41&0.06&0.01&0.00&0.00&0.00&17.50&1.63&0.16&0.02&0.00&0.00&0.00&0.00\\
2&3.33&38.17&5.09&0.60&0.10&0.02&0.01&0.01&1.22&15.10&1.71&0.24&0.03&0.00&0.00&0.00\\
3&0.26&3.97&36.08&5.17&0.61&0.11&0.02&0.01&0.12&1.47&14.04&1.56&0.23&0.02&0.00&0.00\\
4&0.07&0.32&4.40&35.14&5.09&0.54&0.09&0.02&0.04&0.19&1.55&13.01&1.48&0.21&0.02&0.00\\
5&0.04&0.09&0.35&4.36&34.36&4.96&0.46&0.05&0.02&0.07&0.17&1.61&12.40&1.38&0.17&0.01\\
6&0.02&0.05&0.11&0.35&4.25&34.21&4.82&0.28&0.01&0.03&0.05&0.19&1.53&11.89&1.27&0.06\\
7&0.02&0.03&0.05&0.11&0.33&3.94&34.54&2.69&0.01&0.01&0.02&0.07&0.17&1.45&11.12&0.65\\
8&0.02&0.04&0.03&0.08&0.17&0.40&3.68&40.71&0.00&0.01&0.02&0.03&0.08&0.27&1.50&11.99\\
\hline\hline
\end{tabular}
\end{table*}

\begin{table}[htbp]\centering
\caption{
The efficiency matrices for $D_s^+\to\eta^\prime e^+\nu_e$ averaged over all 14 ST modes, where $\varepsilon_{ij}$ represents the
efficiency in \% for events produced in the $j$-th $q^2$ interval and reconstructed in the $i$-th $q^2$ interval. Efficiencies do not include the branching fractions of $\eta^{(\prime)}$ decays.
}
\label{tab:effmatrixb}
\begin{tabular}{c|ccc|ccc}\hline\hline
\multirow{2}{*}{$\varepsilon_{ij}$}&\multicolumn{3}{c|}{$D_s^+\to\eta^\prime_{\eta\pi^+\pi^-}e^+\nu_e$}&\multicolumn{3}{c}{$D_s^+\to\eta^\prime_{\gamma\rho^0}e^+\nu_e$}\\
&1&2&3&1&2&3\\
\hline
1&18.74&1.53&0.06&22.39&1.65&0.08\\
2&0.77&16.99&1.70&0.75&21.54&1.78\\
3&0.01&0.67&17.47&0.05&0.75&22.20\\
\hline\hline
\end{tabular}
\end{table}

\begin{figure*}[htbp]\centering
\setlength{\abovecaptionskip}{-2pt}
\setlength{\belowcaptionskip}{-3pt}
\includegraphics[width=0.95\textwidth]{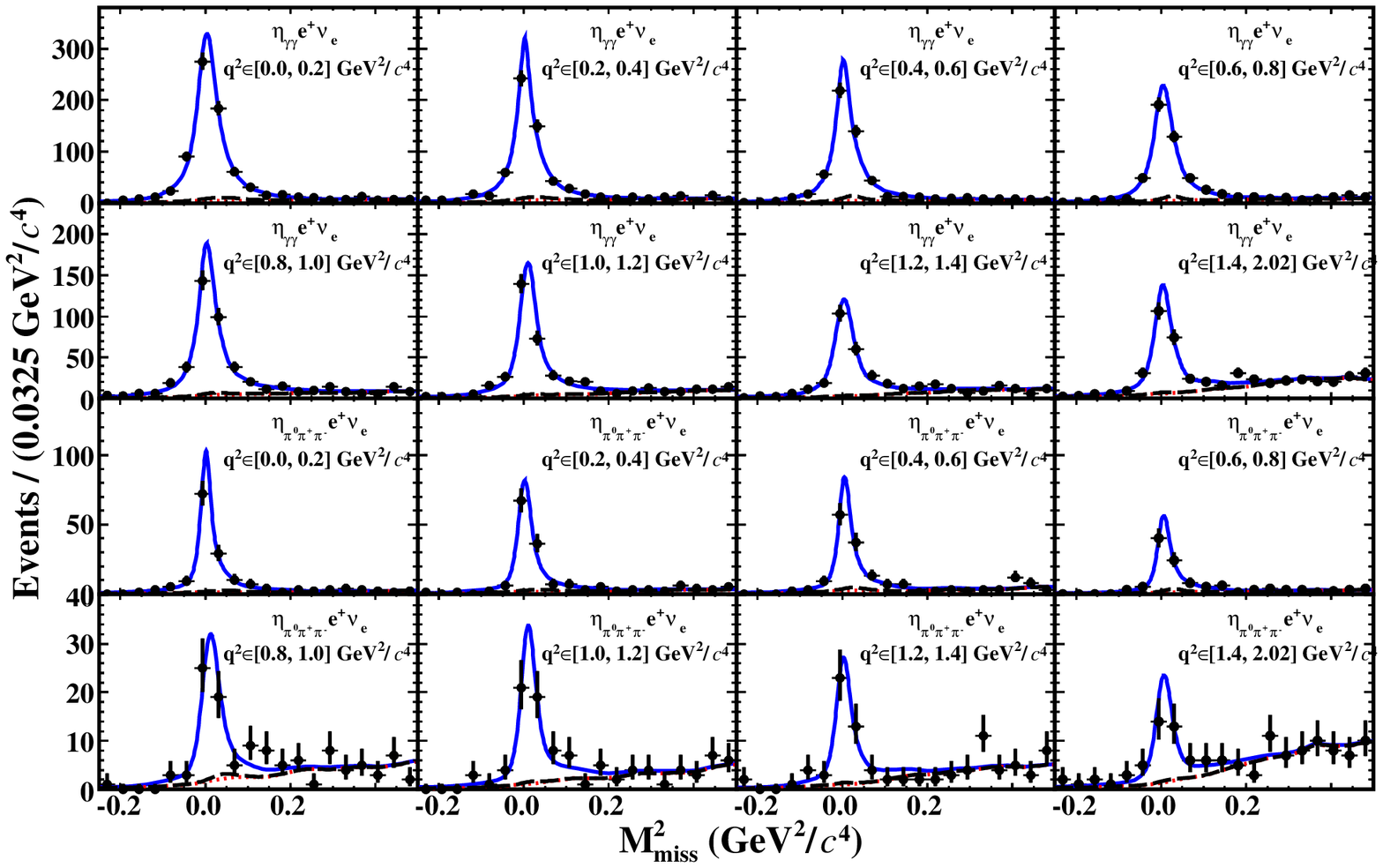}
\caption{Fits to the $M_{\rm miss}^2$ distributions in various reconstructed $q^{2}$ intervals for (top two rows) $D_s^+\to \eta_{\gamma\gamma}e^+\nu_e$ and (bottom two rows) $D_s^+\to \eta_{\pi^0\pi^+\pi^-}e^+\nu_e$.
The points with error bars represent data. The blue solid curves denote the total fits, and the red solid dotted curves show
the fitted combinatorial background contributions. Differences between black dashed and red dotted curves show the backgrounds from $D_s^+\to\eta\pi^+\pi^0$ and $\eta\mu^+\nu_\mu$.
}
\label{fig:fiteachbina}
\end{figure*}

\begin{figure*}[htbp]\centering
\setlength{\abovecaptionskip}{-2pt}
\setlength{\belowcaptionskip}{-3pt}
\includegraphics[width=0.65\textwidth]{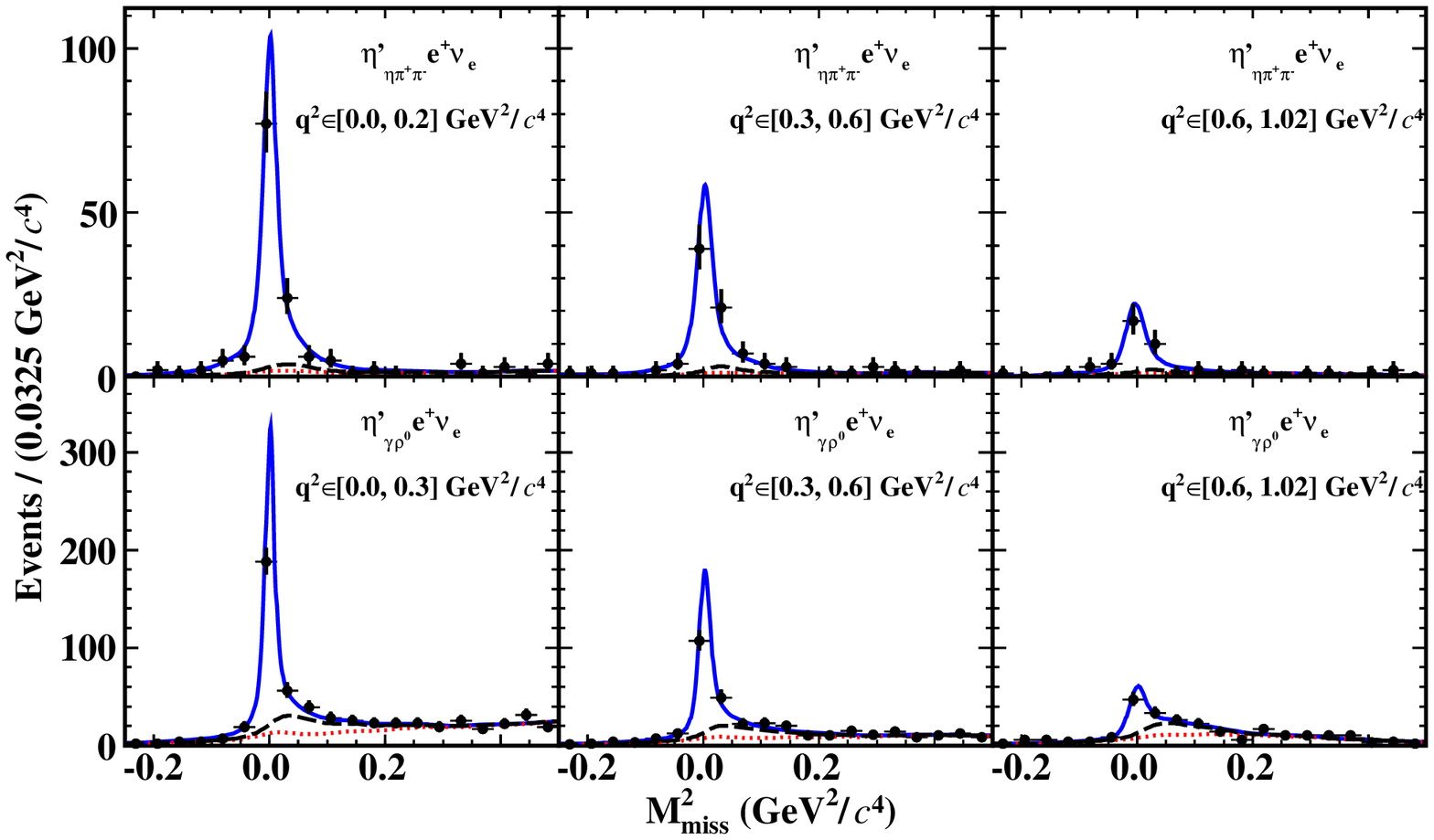}
\caption{Fits to the $M_{\rm miss}^2$ distributions in  various reconstructed $q^{2}$ intervals for (top row) $D_s^+\to \eta^\prime_{\eta\pi^+\pi^-} e^+\nu_e$ and (bottom row) $D_s^+\to \eta^\prime_{\gamma\rho^0} e^+\nu_e$.
The points with error bars represent data. The blue solid curves denote the total fits and the red solid dotted curves show
the fitted combinatorial background contributions. Differences between black dashed and red dotted curves are the backgrounds from $D_s^+\to\eta^{\prime}\pi^+\pi^0$, $\eta^{\prime}\mu^+\nu_\mu$, and $D_s^+ \to \phi(1020)_{\pi^0\pi^+\pi^-} e^+ \nu_e$.
}
\label{fig:fiteachbinb}
\end{figure*}

The number of events observed in each reconstructed $q^{2}$ interval is obtained  
from a fit to the $M^2_{\rm miss}$ distribution of the corresponding events.  Figures~\ref{fig:fiteachbina} and~\ref{fig:fiteachbinb} show the results of the fits to the $M_{\rm miss}^2$ distributions in the reconstructed $q^{2}$ intervals.
Tables~\ref{tab:decayratea} and \ref{tab:decayrateb} summarize the $q^{2}$ ranges, the fitted numbers of observed DT events ($N_{\rm DT}$),
the numbers of generated events ($N_{\rm prd}$) calculated by the weighted efficiency matrix
and the decay rates of $D_s^+\to \eta^{(\prime)}e^+\nu_e$ ($\Delta\Gamma$) in the individual $q^2$ intervals.

\subsection{$\chi^2$ construction and statistical covariance matrices}

To extract the hadronic transition form factor parameters and $|V_{cs}|$, 
the smallest $\chi^{2}$ method is used to fit the partial decay rates of the different signal decays. 
Considering the correlations of the measured partial decay rates ($\Delta\Gamma_i^{\rm msr}$) among different $q^{2}$ intervals, the $\chi^{2}$ is given by
\begin{equation}\label{eq:chi}
	\chi^{2} = \sum_{i,j=1}^{N_{\mathrm{intervals}}}\left(\Delta\Gamma_{i}^{\mathrm{msr}}-\Delta\Gamma_{i}^{\mathrm{th}}\right) C_{ij}^{-1}\left(\Delta\Gamma_{j}^{\mathrm{msr}}-\Delta\Gamma_{j}^{\mathrm{th}}\right),
\end{equation}

	where $\Delta\Gamma_i^{\rm th}$ is the theoretically expected decay rate in channel $i$, $C_{ij}$ is the element of the covariance matrix of the measured partial decay rates and it is given by $C_{ij} = C_{ij}^{\mathrm{stat}}+C_{ij}^{\mathrm{sys}}$.
Here, $C_{ij}^{\mathrm{stat}}$ and $C_{ij}^{\mathrm{sys}}$ are elements of the
statistical and systematic covariance matrices, respectively.
The elements of the statistical covariance matrix are defined as
\begin{equation}
	C_{ij}^{\rm stat} =\left (\frac{1}{\tau_{D^{+}_s}N_{\mathrm{tag}}}\right)^{2}\sum_{\alpha}\varepsilon_{i\alpha}^{-1}\varepsilon_{j\alpha}^{-1}\left(\sigma\left(N_{\mathrm{DT}}^{\alpha}\right)\right)^{2}.
\end{equation}
	Tables~\ref{tab:cova} and \ref{tab:covb} give the elements of the statistical correction density matrices for $D_s^+\to \eta_{\gamma\gamma}e^+\nu_e$, $D_s^+\to \eta_{\pi^0\pi^+\pi^-}e^+\nu_e$, $D_s^+\to \eta^\prime_{\eta\pi^+\pi^-} e^+\nu_e$, and $D_s^+\to \eta^\prime_{\gamma\rho^0} e^+\nu_e$, respectively.

\begin{table*}[htbp]\centering
\caption{The partial decay rates of $D_s^+\to\eta e^+\nu_e$ in various $q^{2}$ intervals. Numbers in the parentheses are the statistical uncertainties. }
\label{tab:decayratea}
\scalebox{0.93}{
\begin{tabular}{cccccccccc}\hline\hline
\multicolumn{2}{c}{$i$}&1&2&3&4&5&6&7&8\\
\multicolumn{2}{c}{$q^2$ $(\mathrm{GeV}^{2}/c^{4})$}&($m_e^2,\,0.2$)&($0.2,\,0.4$)&($0.4,\,0.6$)&($0.6,\,0.8$)&($0.8,\,1.0$)&($1.0,\,1.2$)&($1.2,\,1.4$)&($1.4,\,2.02$)\\
\hline
\multirow{3}{*}{$D_s^+\to\eta_{\gamma\gamma} e^+\nu_e$}&$N_{\mathrm{obs}}^i$&681(28)&550(26)&497(25)&455(24)&371(21)&316(20)&252(18)&260(19)\\
&$N_{\mathrm{prd}}^i$&3637(167)&2929(179)&2740(181)&2578(180)&2092(167)&1826(154)&1497(137)&1446(118)\\
&$\Delta\Gamma^i_{\rm msr}$ $(\mathrm{ns^{-1}})$&8.83(40)&7.11(43)&6.66(44)&6.26(44)&5.08(41)&4.43(37)&3.64(33)&3.51(29)\\
\hline
\multirow{3}{*}{$D_s^+\to\eta_{\pi^0\pi^+\pi^-} e^+\nu_e$}&$N_{\mathrm{DT}}^i$&132(12)&123(12)&120(12)&83(10)&58(09)&57(09)&47(08)&42(08)\\
&$N_{\mathrm{prd}}^i$&3019(316)&2951(365)&3153(400)&2169(359)&1486(345)&1688(337)&1500(317)&1298(303)\\
&$\Delta\Gamma^i_{\rm msr}$ $(\mathrm{ns^{-1}})$&7.33(77)&7.17(89)&7.66(97)&5.27(87)&3.61(84)&4.10(82)&3.64(77)&3.15(73)\\
\hline\hline
\end{tabular}
}
\end{table*}

\begin{table*}[htbp]\centering
\caption{
The partial decay rates of $D_s^+\to\eta^\prime e^+\nu_e$ in various $q^{2}$ intervals. Numbers in the parentheses are the statistical uncertainties. }
\label{tab:decayrateb}

\scalebox{0.95}{
\begin{tabular}{ccccc}\hline\hline
\multicolumn{2}{c}{$i$}&1&2&3\\
\multicolumn{2}{c}{$q^2$ $(\mathrm{GeV}^{2}/c^{4})$}&($m_e^2,\,0.3$)&($0.3,\,0.6$)&($0.6,\,1.02$)\\
\hline
\multirow{3}{*}{$D_s^+\to\eta^\prime_{\eta_{\gamma\gamma}\pi^+\pi^-} e^+\nu_e$}&$N_{\mathrm{obs}}^i$&116(12)&72(09)&32(07)\\
&$N_{\mathrm{prd}}^i$&3528(371)&2258(331)&991(226)\\
&$\Delta\Gamma^i_{\rm msr}$ $(\mathrm{ns^{-1}})$&8.57(90)&5.48(80)&2.41(55)\\
\hline
\multirow{3}{*}{$D_s^+\to\eta^\prime_{\gamma\rho^0} e^+\nu_e$}&$N_{\mathrm{DT}}^i$&237(19)&157(16)&63(14)\\
&$N_{\mathrm{prd}}^i$&3411(286)&2281(251)&879(217)\\
&$\Delta\Gamma^i_{\rm msr}$ $(\mathrm{ns^{-1}})$&8.28(69)&5.54(61)&2.14(53)\\
\hline\hline
\end{tabular}
}

\end{table*}

\subsection{Systematic uncertainties}

Several sources of systematic uncertainties are discussed below.  

\paragraph{$D^{+}_s$ lifetime}

	The uncertainties associated with the $D^{+}_s$ lifetime are fully correlated across the $q^2$ intervals. The element of the related systematic covariance matrix is calculated by
\begin{equation}
C_{ij}^{\mathrm{sys}}\left(\tau_{D^+_s}\right)=\sigma\left(\Delta\Gamma_{i}\right)\sigma\left(\Delta\Gamma_{j}\right),
\end{equation}
where $\sigma(\Delta\Gamma_{i})=\sigma \tau_{D^+_s}\cdot\Delta\Gamma_{i}$ and $\sigma \tau_{D^+_s}$ is the uncertainty of the $D^{+}_s$ lifetime~\cite{PDG2022}.

\paragraph{MC statistics}

	Systematic efficiency uncertainties in and correlations between the $q^{2}$ intervals due to the limited MC size are calculated by
\begin{equation}
C_{ij}^{\mathrm{sys}}=\left(\frac{1}{\tau_{D^+_s}N_{\mathrm{tag}}}\right)^{2}\sum_{\alpha\beta}N_{\mathrm{DT}}^{\alpha}N_{\mathrm{DT}}^{\beta}\mathrm{Cov}\left(\varepsilon_{i\alpha}^{-1},\varepsilon_{j\beta}^{-1}\right),
\end{equation}
where the covariances of the inverse efficiency matrix elements are given by~\cite{Lefebvre:1999yu}
\begin{equation}
\mathrm{Cov}\left(\varepsilon_{i\alpha}^{-1},\varepsilon_{j\beta}^{-1}\right)=\sum\limits_{mn}\left(\varepsilon_{im}^{-1}\varepsilon_{j m}^{-1}\right)\left [\sigma\left (\epsilon_{mn}\right )\right]^2\left (\varepsilon_{\alpha n}^{-1}\varepsilon_{\beta n}^{-1}\right).
\end{equation}

\paragraph{Hadronic transition form factor}

	Systematic uncertainties associated with the hadronic transition form factor used to generate signal MC events are estimated by re-weighting the signal MC events so that the $q^{2}$ spectrum agrees with the measured spectrum. For each signal MC event, the weight $\omega$ is given by
\begin{equation}
\omega = \frac{\left |f_{+}^{\eta^{(\prime)}~\mathrm{measured}}(q^{2})\right |^{2}\int_{q^{2}_{\mathrm{min}}}^{q^{2}_{\mathrm{max}}}\frac{\mathrm{d}\Gamma^{\mathrm{default}}}{\mathrm{d}q^{2}}\mathrm{d}q^{2}}{\left |f_{+}^{\eta^{(\prime)}\mathrm{default}}(q^{2})\right |^{2}\int_{q^{2}_{\mathrm{min}}}^{q^{2}_{\mathrm{max}}}\frac{\mathrm{d}\Gamma^{\mathrm{measured}}}{\mathrm{d}q^{2}}\mathrm{d}q^{2}},
\end{equation}
where $f_{+}^{\eta^{(\prime)}\mathrm{default}}(q^{2})$ is the default hadronic transition form factor used to generate the signal MC events. The default hadronic transition form factor uses the modified pole model with the parameter $\alpha = 0.25$ and $f_{+}(0)=1.0$. The $f_{+}^{\eta^{(\prime)}~\mathrm{measured}}(q^{2})$ is the measured hadronic transition form factor for $D_s^+\to\eta^{(\prime)} e^+\nu_e$ using the 2-Par series expansion with parameters obtained from the fit with the statistical covariance matrix.

The partial decay rates are then calculated  in different $q^{2}$ intervals with the newly weighted efficiency matrix. The element of the covariance matrix is defined as
\begin{equation}
C_{ij}^{\mathrm{sys}}\left (\mathrm{FF}\right )=\delta\left(\Delta\Gamma_{i}\right)\delta\left(\Delta\Gamma_{j}\right),
\end{equation}
	where $\delta(\Delta\Gamma_{i})$ denotes the change of the partial decay rate in the $i$-th $q^{2}$ interval.

\paragraph{Tracking, PID, and $\gamma, \eta, \pi^0$ reconstruction}

The systematic uncertainties associated with the $e^+$ tracking and PID efficiencies, pion tracking and PID efficiencies, and $\gamma, \eta, \pi^0$ reconstruction are estimated by varying the corresponding correction factors for efficiencies within $\pm 1\sigma$. Using the new efficiency matrix, the element of the corresponding systematic covariance matrix is calculated by
\begin{equation}
C_{ij}^{\mathrm{sys}}\left (\mathrm{Tracking,\,PID,\,\gamma/\eta/\pi^0 rec.}\right)=\delta\left(\Delta\Gamma_{i}\right)\delta\left(\Delta\Gamma_{j}\right),
\end{equation}
where $\delta(\Delta\Gamma_{i})$ denotes the change of the partial decay rate in the $i$-th $q^{2}$ interval.

\paragraph{$M_{\rm miss}^2$ fit}

The systematic covariance matrix arising from the uncertainty in the $M_{\rm miss}^2$ fit has elements
\begin{equation}
C_{ij}^{\mathrm{sys}}\left(M_{\rm miss}^2\,\,\mathrm{fit}\right)=\left(\frac{1}{\tau_{D^+_s}N_{\mathrm{tag}}}\right)^{2}\sum_{\alpha}\varepsilon_{i\alpha}^{-1}\varepsilon_{j\alpha}^{-1}\left(\sigma_{\alpha}^{\mathrm{fit}}\right)^{2},
\end{equation}
where $\sigma_{\alpha}^{\mathrm{fit}}$ is the systematic uncertainty of the number of signal events observed in the interval $\alpha$ obtained by varying the background shape in the $M_{\rm miss}^2$ fit.

\paragraph{Remaining uncertainties}

The remaining uncertainties are assumed to be  fully correlated across $q^{2}$ intervals and the element of the corresponding systematic covariance matrix is calculated by
\begin{equation}
C_{ij}^{\mathrm{sys}}=\sigma\left(\Delta\Gamma_{i}\right)\sigma\left(\Delta\Gamma_{j}\right),
\end{equation}
where $\sigma(\Delta\Gamma_{i})=\sigma_{\rm sys}\cdot\Delta\Gamma_{i}$ and $\sigma_{\rm sys}$ is the corresponding uncertainty reported in Table~\ref{sys}.

Tables~\ref{table:sysffa} and \ref{table:sysffb} give the systematic uncertainties for all sources in the different $q^2$ intervals, and
Tables~\ref{tab:cova} and \ref{tab:covb} give the elements of the systematic covariance density matrices for $D_s^+\to\eta e^+\nu_e$ and $D_s^+\to\eta^\prime e^+\nu_e$, respectively.

\setlength{\tabcolsep}{3pt}

\begin{table*}
\centering
\caption{Systematic uncertainties (in \%) of the measured decay rates of $D_s^+\to\eta e^+\nu_e$ in various $q^{2}$ intervals. }
\label{table:sysffa}
\scalebox{0.9}{
\begin{tabular}{l|cccccccc|cccccccc}
\hline\hline

&\multicolumn{8}{c|}{$D_s^+\to\eta_{\gamma\gamma}e^+\nu_e$}&\multicolumn{8}{c}{$D_s^+\to\eta_{\pi^0\pi^+\pi^-}e^+\nu_e$}\\
&1&2&3&4&5&6&7&8&1&2&3&4&5&6&7&8\\
\hline
ST $D_s^-$ yields&0.50&0.50&0.50&0.50&0.50&0.50&0.50&0.50&0.50&0.50&0.50&0.50&0.50&0.50&0.50&0.50\\
$D_s^+$ lifetime&0.80&0.80&0.80&0.80&0.80&0.80&0.80&0.80&0.80&0.80&0.80&0.80&0.80&0.80&0.80&0.80\\
$\pi^0/\eta$ reconstruction&1.84&1.59&1.32&1.08&0.96&0.90&0.89&0.87&0.94&1.00&1.06&1.16&1.27&1.37&1.59&1.90\\$\pi^\pm$ tracking&--&--&--&--&--&--&--&--&0.95&1.01&1.09&1.17&1.27&1.38&1.51&1.74\\
$\pi^\pm$ PID&--&--&--&--&--&--&--&--&0.16&0.17&0.18&0.21&0.24&0.26&0.31&0.39\\
$e^+$ tracking&0.50&0.50&0.50&0.50&0.50&0.50&0.50&0.50&0.50&0.50&0.50&0.50&0.50&0.50&0.50&0.50\\
$e^+$ PID&0.08&0.09&0.09&0.10&0.11&0.10&0.12&0.12&0.06&0.08&0.09&0.11&0.09&0.12&0.11&0.11\\
Transition $\gamma(\pi^0)$ reconstruction&1.00&1.00&1.00&1.00&1.00&1.00&1.00&1.00&1.00&1.00&1.00&1.00&1.00&1.00&1.00&1.00\\
Smallest $|\Delta E|$&1.00&1.00&1.00&1.00&1.00&1.00&1.00&1.00&1.00&1.00&1.00&1.00&1.00&1.00&1.00&1.00\\
Peaking background&0.40&0.40&0.40&0.40&0.40&0.40&0.40&0.40&0.40&0.40&0.40&0.40&0.40&0.40&0.40&0.40\\
Hadronic transition form factors&0.17&1.29&2.22&0.85&0.07&0.81&0.79&0.19&0.27&0.93&5.06&4.21&6.96&1.56&1.84&8.22\\

$\eta_{\pi^0\pi^+\pi^-}$ selection&--&--&--&--&--&--&--&--&0.10&0.10&0.10&0.10&0.10&0.10&0.10&0.10\\
Tag bias&0.40&0.40&0.40&0.40&0.40&0.40&0.40&0.40&0.10&0.10&0.10&0.10&0.10&0.10&0.10&0.10\\

$E_{\rm extra,~\gamma}^{\rm max}$ and $N^{\rm char}_{\rm extra} $ requirements&0.70&0.70&0.70&0.70&0.70&0.70&0.70&0.70&2.00&2.00&2.00&2.00&2.00&2.00&2.00&2.00\\

$M_{\rm miss}^{2}$ fit&0.10&0.37&0.24&0.48&0.69&3.05&0.85&0.38&0.40&0.11&0.38&0.57&2.01&1.35&1.89&4.00\\
MC statistics&0.92&1.14&1.26&1.35&1.51&1.65&1.80&1.49&1.14&1.37&1.46&1.74&2.06&2.09&2.42&2.19\\
Quoted branching fractions&0.50&0.50&0.50&0.50&0.50&0.50&0.50&0.50&1.10&1.10&1.10&1.10&1.10&1.10&1.10&1.10\\

\hline
Total&2.91&3.14&3.54&2.85&2.81&4.20&3.10&2.72&3.44&3.65&6.23&5.69&8.28&4.59&5.12&10.18\\
\hline
\hline
\end{tabular}
}
\end{table*}

\begin{table*}[htbp]
\centering
\caption{Systematic uncertainties (\%) of the measured decay rates of $D_s^+\to \eta^\prime e^+\nu_e$ in various $q^{2}$ intervals.}
\label{table:sysffb}
\setlength{\tabcolsep}{2mm}{
\begin{tabular}{l|ccc|ccc}
 \hline\hline
 
&\multicolumn{3}{c|}{$D_s^+\to\eta^\prime_{\eta\pi^+\pi^-}e^+\nu_e$}&\multicolumn{3}{c}{$D_s^+\to\eta^\prime_{\gamma\rho^0}e^+\nu_e$}\\
&1&2&3&1&2&3\\
   \hline
ST $D_s^-$ yields&0.50&0.50&0.50&0.50&0.50&0.50\\
$D_s^+$ lifetime&0.80&0.80&0.80&0.80&0.80&0.80\\
$\pi^0/\eta$ reconstruction&0.76&0.81&0.96&--&--&--\\
$\pi^\pm$ tracking&1.65&1.79&1.96&0.67&0.70&0.74\\
$\pi^\pm$ PID&0.35&0.40&0.47&0.11&0.09&0.08\\
$e^+$ tracking&0.50&0.50&0.50&0.50&0.50&0.50\\
$e^+$ PID&0.11&0.12&0.14&0.11&0.11&0.13\\
Transition $\gamma(\pi^0)$ reconstruction&1.00&1.00&1.00&1.00&1.00&1.00\\
Smallest $|\Delta E|$&1.00&1.00&1.00&1.00&1.00&1.00\\
Peaking background&1.00&1.00&1.00&1.00&1.00&1.00\\
Hadronic transition form factors&1.58&2.75&0.76&0.26&1.69&0.90\\

$\eta^{\prime}$ selection&0.10&0.10&0.10&1.40&1.40&1.40\\
Tag bias&0.10&0.10&0.10&0.10&0.10&0.10\\
$\chi^{2}$ requirement&--&--&--&1.50&1.50&1.50\\

$E_{\rm extra,~\gamma}^{\rm max}$ and $N^{\rm char}_{\rm extra}$ requirements&2.00&2.00&2.00&1.10&1.10&1.10\\
$M_{\rm miss}^{2}$ fit&0.52&0.50&1.27&0.30&0.75&9.18\\

MC statistics&0.35&0.48&0.72&0.71&0.91&1.38\\
Quoted branching fractions&1.30&1.30&1.30&1.40&1.40&1.40\\

\hline
Total&4.02&4.69&4.19&3.56&4.04&9.95\\
\hline
\hline
\end{tabular}
}
\end{table*}

\begin{table*}[htp]\centering
\caption{Statistical and systematic uncertainty density matrices for the measured partial decay rates of $D_s^+\to\eta e^+\nu_e$ in different $q^2$ intervals.}
\label{tab:cova}
\scalebox{0.9}{
\begin{tabular}{ccccccccccccccccc}\hline\hline
\multicolumn{17}{c}{Statistical correlation matrix}\\
\multirow{2}{*}{$\rho_{ij}^{\rm stat}$}&\multicolumn{8}{c}{$D_s^+\to\eta_{\gamma\gamma}e^+\nu_e$}&\multicolumn{8}{c}{$D_s^+\to\eta_{\pi^0\pi^+\pi^-}e^+\nu_e$}\\
&1&2&3&4&5&6&7&8&1&2&3&4&5&6&7&8\\
\hline
1	&1.000	&-0.187	&0.019	&-0.003	&0.000	&0.000	&0.000	&0.000	&0.000	&0.000	&0.000	&0.000	&0.000	&0.000&0.000	&0.000	\\
2	&&1.000	&-0.236	&0.025	&-0.004	&0.000	&0.000	&-0.001	&0.000	&0.000	&0.000	&0.000	&0.000	&0.000	&0.000&0.000	\\
3	&&&1.000	&-0.259	&0.029	&-0.005	&0.000	&0.000	&0.000	&0.000	&0.000	&0.000	&0.000	&0.000	&0.000&0.000	\\
4	&&&&1.000	&-0.262	&0.030	&-0.006	&-0.001	&0.000	&0.000	&0.000	&0.000	&0.000	&0.000	&0.000	&0.000\\
5	&&&&&1.000	&-0.259	&0.029	&-0.005	&0.000	&0.000	&0.000	&0.000	&0.000	&0.000	&0.000	&0.000	\\
6	&&&&&&1.000	&-0.247	&0.017	&0.000	&0.000	&0.000	&0.000	&0.000	&0.000	&0.000	&0.000	\\
7	&&&&&&&1.000	&-0.168	&0.000	&0.000	&0.000	&0.000	&0.000	&0.000	&0.000	&0.000	\\
8	&&&&&&&&1.000	&0.000	&0.000	&0.000	&0.000	&0.000	&0.000	&0.000	&0.000	\\
1	&&&&&&&&&1.000	&-0.174	&0.012	&-0.002	&-0.001	&0.000	&-0.001	&0.000	\\
2	&&&&&&&&&&1.000	&-0.212	&0.011	&-0.003	&-0.001	&-0.001	&0.000	\\
3	&&&&&&&&&&&1.000	&-0.223	&0.012	&-0.001	&-0.001	&-0.001	\\
4	&&&&&&&&&&&&1.000	&-0.234	&0.012	&-0.005	&-0.001	\\
5	&&&&&&&&&&&&&1.000	&-0.230	&0.016	&-0.003	\\
6	&&&&&&&&&&&&&&1.000	&-0.231	&0.008	\\
7	&&&&&&&&&&&&&&&1.000	&-0.180	\\
8	&&&&&&&&&&&&&&&&1.000	\\

\hline
\multicolumn{17}{c}{Systematic correlation matrix}\\
\multirow{2}{*}{$\rho_{ij}^{\rm syst}$}&\multicolumn{8}{c}{$D_s^+\to\eta_{\gamma\gamma}e^+\nu_e$}&\multicolumn{8}{c}{$D_s^+\to\eta_{\pi^0\pi^+\pi^-}e^+\nu_e$}\\
&1&2&3&4&5&6&7&8&1&2&3&4&5&6&7&8\\
\hline
1	&1.000	&0.738	&0.609	&0.734	&0.738	&0.468	&0.672	&0.730	&0.497	&0.469	&0.243	&0.371	&0.184	&0.417&0.397	&0.277	\\
2	&&1.000	&0.790	&0.805	&0.673	&0.270	&0.486	&0.673	&0.476	&0.532	&0.590	&-0.016	&0.551	&0.522	&0.511&-0.134	\\
3	&&&1.000	&0.701	&0.578	&0.256	&0.337	&0.610	&0.422	&0.518	&0.724	&-0.225	&0.698	&0.530	&0.523&-0.345	\\
4	&&&&1.000	&0.633	&0.262	&0.548	&0.683	&0.463	&0.497	&0.493	&0.061	&0.448	&0.468	&0.451	&-0.055\\
5	&&&&&1.000	&0.234	&0.652	&0.674	&0.437	&0.422	&0.267	&0.259	&0.214	&0.368	&0.345	&0.161	\\
6	&&&&&&1.000	&0.230	&0.458	&0.272	&0.225	&0.006	&0.325	&-0.035	&0.170	&0.151	&0.274	\\
7	&&&&&&&1.000	&0.536	&0.368	&0.306	&0.012	&0.436	&-0.043	&0.232	&0.206	&0.365	\\
8	&&&&&&&&1.000	&0.446	&0.438	&0.306	&0.228	&0.254	&0.385	&0.362	&0.124	\\
1	&&&&&&&&&1.000	&0.829	&0.565	&0.495	&0.472	&0.767	&0.712	&0.313	\\
2	&&&&&&&&&&1.000	&0.660	&0.340	&0.584	&0.763	&0.721	&0.129	\\
3	&&&&&&&&&&&1.000	&-0.302	&0.917	&0.699	&0.686	&-0.437	\\
4	&&&&&&&&&&&&1.000	&-0.376	&0.215	&0.185	&0.853	\\
5	&&&&&&&&&&&&&1.000	&0.644	&0.688	&-0.426	\\
6	&&&&&&&&&&&&&&1.000	&0.715	&0.117	\\
7	&&&&&&&&&&&&&&&1.000	&0.110	\\
8	&&&&&&&&&&&&&&&&1.000	\\

\hline\hline
\end{tabular}
}
\end{table*}

\begin{table}[htp]\centering
\caption{Statistical and systematic uncertainty density matrices for the measured partial decay rates of  $D_s^+\to\eta^\prime e^+\nu_e$ in different
$q^2$ intervals.}
\label{tab:covb}
\begin{tabular}{ccccccc}\hline\hline
\multicolumn{7}{c}{Statistical correlation matrix}\\
\multirow{2}{*}{$\rho_{ij}^{\rm stat}$}&\multicolumn{3}{c}{$D_s^+\to\eta^\prime_{\eta\pi^+\pi^-}e^+\nu_e$}&\multicolumn{3}{c}{$D_s^+\to\eta^\prime_{\gamma\rho^0}e^+\nu_e$}\\
&1&2&3&1&2&3\\
\hline
1	&1.000	&-0.123	&0.009	&0.000	&0.000	&0.000	\\
2	&&1.000	&-0.124	&0.000	&0.000	&0.000	\\
3	&&&1.000	&0.000	&0.000	&0.000	\\
1	&&&&1.000	&-0.104	&0.003	\\
2	&&&&&1.000	&-0.110	\\
3	&&&&&&1.000	\\
\hline
\multicolumn{7}{c}{Systematic correlation matrix}\\
\multirow{2}{*}{$\rho_{ij}^{\rm syst}$}&\multicolumn{3}{c}{$D_s^+\to\eta^\prime_{\eta\pi^+\pi^-}e^+\nu_e$}&\multicolumn{3}{c}{$D_s^+\to\eta^\prime_{\gamma\rho^0}e^+\nu_e$}\\
&1&2&3&1&2&3\\
\hline
1	&1.000	&0.508	&0.924	&0.348	&0.121	&0.080	\\
2	&&1.000	&0.648	&0.236	&0.492	&0.153	\\
3	&&&1.000	&0.332	&0.204	&0.098	\\
1	&&&&1.000	&0.818	&0.311	\\
2	&&&&&1.000	&0.258	\\
3	&&&&&&1.000	\\

\hline\hline
\end{tabular}
\end{table}

\subsection{Results based on individual fits}

\begin{figure*}[htp]\centering
\setlength{\abovecaptionskip}{-2pt}
\setlength{\belowcaptionskip}{-3pt}
\includegraphics[width=0.95\textwidth]{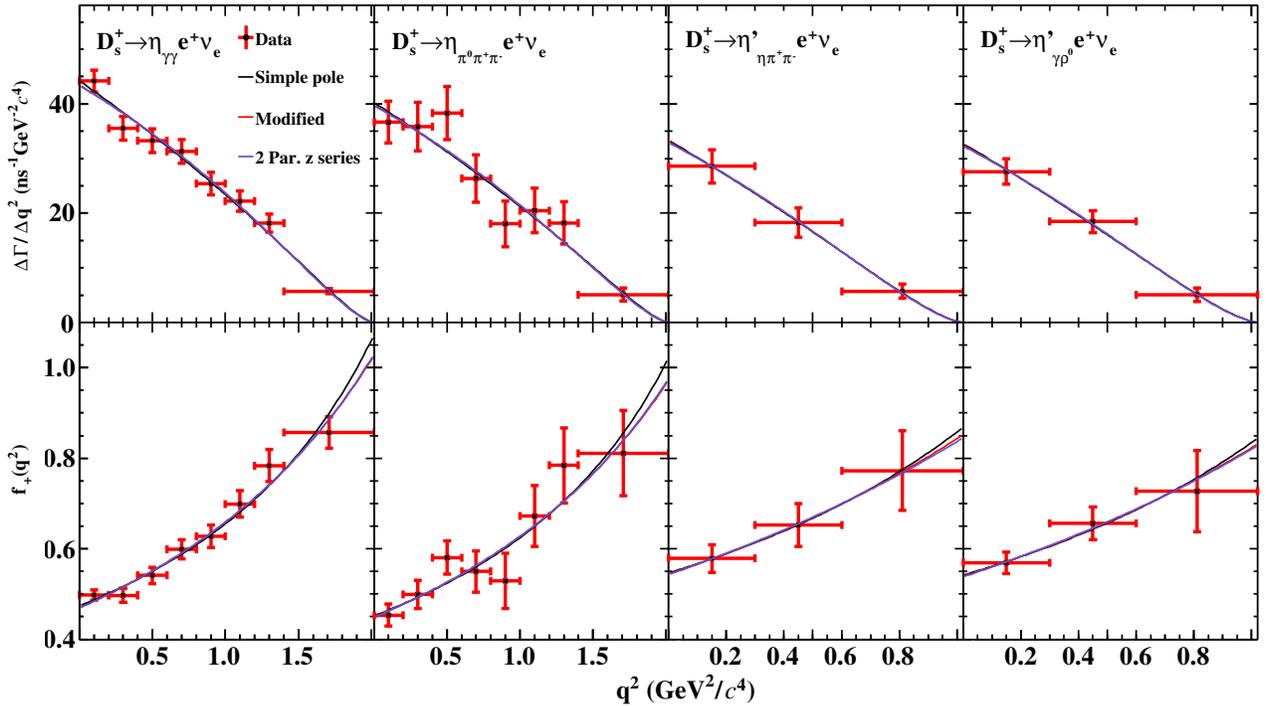}
\caption{(Top row) Individual fits to the differential decay rates of $D^+_s\to \eta e^+\nu_e$ and $D^+_s\to \eta^\prime e^+\nu_e$
and (bottom row)  projection on the hadronic transition form factors as a function of $q^2$. The points with error bars are (top row) the measured differential decay rates and (bottom row) the hadronic transition form factors. The black, red, and blue curves are the form factors parameterized by simple pole model, modified pole model, and 2-Par series expansion, respectively.
}
\label{fig:fitdecayrate}
\end{figure*}

For each semileptonic decay, the product $f^{\eta^{(\prime)}}_+(0)|V_{cs}|$ 
and one of the parameters $M_{\rm pole}$, $\alpha$, or $r_1$ are
determined by constructing and minimizing the $\chi^2$ defined in Eq.~\ref{eq:chi}.
The covariance matrices used in these fits are shown in Tables~\ref{tab:cova} and~\ref{tab:covb}. 
Figure \ref{fig:fitdecayrate} shows individual fits to the differential decay rates of $D^+_s\to \eta e^+\nu_e$ and $D^+_s\to \eta^\prime e^+\nu_e$ and (second row) the hadronic transition form factors as a function of $q^2$. The results obtained from individual fits are listed in Table~\ref{table:parameter}.

\begin{table*}[htp]
\centering
\caption{The parameters obtained from individual fits to the partial decay rates of $D^+_s\to \eta e^+\nu_e$ or
$D^+_s\to \eta^\prime e^+\nu_e$. The first and second uncertainties are statistical and systematic, respectively.
$N_{\rm d.o.f}$ is the number of degrees of freedom.
\label{table:parameter}
}
\scalebox{0.83}{
  \begin{tabular}{l|ccc|ccc|ccc}\hline\hline
     &\multicolumn{3}{c}{Simple pole}  &  \multicolumn{3}{c}{Modified pole}  &  \multicolumn{3}{c}{Series 2-Par} \\
  Decay &  $f_+^{\eta^{(\prime)}}(0)|V_{cs}|$           &  $M_{\rm pole}$        &   $\chi^2/{N_{\rm d.o.f}}$& $f_+^{\eta^{(\prime)}}(0)|V_{cs}|$           &  $\alpha$              &   $\chi^2/{N_{\rm d.o.f}}$& $f_+^{\eta^{(\prime)}}(0)|V_{cs}|$           & $r_1$                  &   $\chi^2/{N_{\rm d.o.f}}$\\\hline
   ${ \eta_{\gamma\gamma} e^{+} \nu_e}$&$0.4613(67)(65)$    &$1.90(04)(02)$    &5.3/6 &$0.4568(75)(67)$    &  $0.37(09)(03)$  &    4.9/6&$0.4570(78)(68)$    &$-2.59(45)(15)$      &4.9/6\\
      ${ \eta_{\pi^0\pi^+\pi^-} e^{+} \nu_e}$&$0.440(15)(08)$    &$1.90(11)(03)$    &5.0/6 &$0.437(16)(08)$    &  $0.35(21)(06)$  &    5.0/6&$0.437(17)(08)$    &$-2.5(11)(03)$      &5.0/6\\

   $\eta^{\prime}_{\eta\pi^+\pi^-} e^{+} \nu_e$&$0.532(34)(10) $   &$ 1.66(31)(04)$   &0.0/1 &$0.530(36)(10)$    &  $ 0.75(75)(10)$ &    0.0/1 &  $0.529(39)(11)$  & $-6.5(60)(09) $   &0.0/1\\
      $\eta^{\prime}_{\gamma\rho^0} e^{+} \nu_e$&$0.527(28)(10) $   &$ 1.69(30)(07)$   &0.2/1 &$0.525(29)(11)$    &  $ 0.69(66)(18)$ &    0.2/1 &  $0.524(31)(11)$  & $-6.1(53)(14) $   &0.1/1\\
   \hline\hline
    \end{tabular}
}
\end{table*}

\subsection{Results based on simultaneous fits}

Since the results for the hadronic transition form factors are consistent with each other,
simultaneous fits to the differential decay rates of $D_s^+\to \eta e^+\nu_e$ and $D_s^+\to \eta^\prime e^+\nu_e$ are performed to improve the statistical precision.

\begin{figure*}[htbp]\centering
\setlength{\abovecaptionskip}{2pt}
\setlength{\belowcaptionskip}{-3pt}
\includegraphics[width=0.75\textwidth]{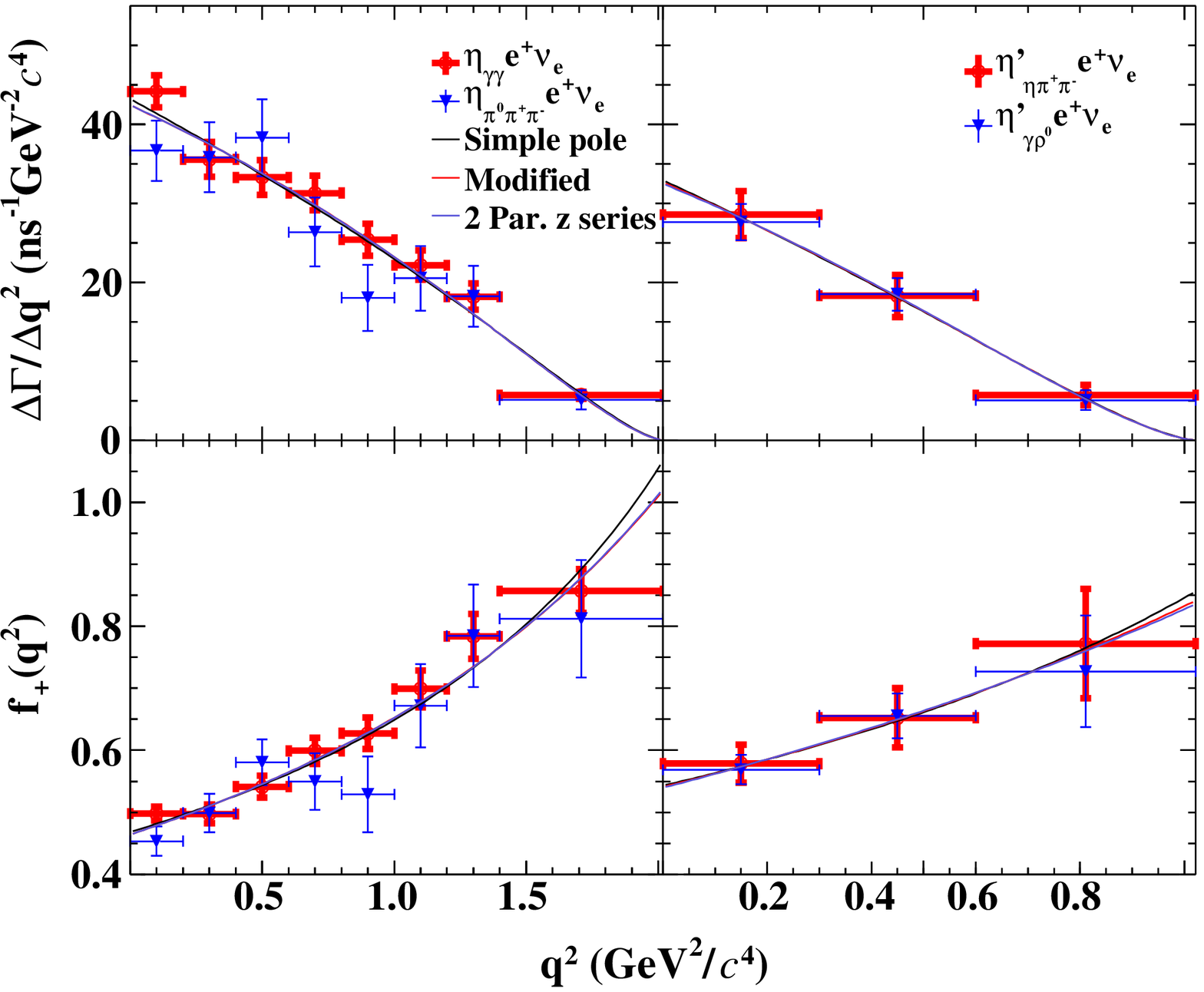}
\caption{(Top row) Simultaneous fits to the differential decay rates of (left) $D^+_s\to \eta_{\gamma\gamma} e^+\nu_e$ and  $D^+_s\to \eta_{\pi^0\pi^+\pi^-} e^+\nu_e$ and (right) $D^+_s\to \eta^\prime_{\eta\pi^+\pi^-} e^+\nu_e$ and  $D^+_s\to \eta^\prime_{\rho^0\gamma} e^+\nu_e$,
and (bottom row)  projection on the hadronic transition form factors as function of $q^2$. The red circles and blue triangles with error bars are (top row) the measured differential decay rates for two $\eta^{(\prime)}$ channels and (bottom row) the hadronic transition form factors. The black, red, and blue curves are the form factors parameterized by simple pole model, modified pole model, and  2-Par series expansion, respectively.
\label{fig:fitdecayrate_combine}}
\end{figure*}

The values of $\Delta\Gamma^i_{\rm msr}$ measured by the two $\eta^{(\prime)}$ sub-decays are fitted
simultaneously, with results shown in Fig.~\ref{fig:fitdecayrate_combine}. In the fits, the
$\Delta\Gamma^i_{\rm msr}$ becomes a vector of length $2m$ and $C_{ij}$ becomes a $2m\times2m$ covariance matrix. 
The uncorrelated and correlated systematic uncertainties are the same as shown in Table~\ref{sys}. 

For fully correlated systematic uncertainties, the matrices are constructed in the same way as done for the individual fits.
For the uncorrelated systematic uncertainties, the matrix takes the form

$$C_{ij}=
\begin{pmatrix}
A&0\\
0&B
\end{pmatrix},
$$
where $A$ and $B$ are the matrices obtained from the individual decays.
Table~\ref{tab:FF_combine} summarizes the fit results obtained from the simultaneous, where the obtained values of
$f_+^{\eta^{(\prime)}}(0)|V_{cs}|$ with different hadronic transition form factor parameterizations are consistent with each
other.

Combining $|V_{cs}| = 0.97349 \pm 0.00016$ from the global fit in the standard model~\cite{PDG2022}
with $f^{\eta^{(\prime)}}_+(0)|V_{cs}|$ extracted from the 2-Par series expansion, we
determine $f^{\eta}_+(0) = 0.4642\pm0.0073_{\rm stat}\pm0.0066_{\rm syst}$ and $f^{\eta^{\prime}}_+(0)=0.540\pm0.025_{\rm stat}\pm0.009_{\rm syst}$. 
Alternatively, we determine $|V_{cs}|$ with $D_s^+\to \eta^{(\prime)} e^+\nu_e$ decays by taking the $f_+^{\eta^{(\prime)}}(0)$ given by theoretical calculations.
 With $f^{\eta}_+(0)=0.495^{+0.030}_{-0.029}$ and $f^{\eta^\prime}_+(0)=0.558^{+0.047}_{-0.045}$ from Ref.~\cite{Duplancic:2015zna}, we obtain $|V_{cs}|_\eta=0.913\pm0.014_{\rm stat}\pm0.013_{\rm syst}{^{+0.055}_{-0.053}}_{\rm theo}$ and $|V_{cs}|_{\eta^\prime}= 0.941\pm0.044_{\rm stat}\pm0.016_{\rm syst}{^{+0.079}_{-0.076}}_{\rm theo}$, where the third uncertainties  originate from the input FFs. These results agree with the measurements of $|V_{cs}|$ using
$D\to \bar K
\ell^+\nu_\ell$~\cite{Ablikim:2015ixa,Ablikim:2015qgt,Ablikim:2018evp,Besson:2009uv,Aubert:2007wg,Widhalm:2006wz}
and $D_s^+ \to \ell^+\nu_\ell$
decays~\cite{Ablikim:2016duz,Zupanc:2013byn,delAmoSanchez:2010jg,Onyisi:2009th,Naik:2009tk}.

\begin{table*}[htp]
\centering
\caption{\small The parameters obtained from simultaneous fits to the partial decay rates of $D^+_s\to \eta e^+\nu_e$ or
$D^+_s\to \eta^\prime e^+\nu_e$. The first and second uncertainties are statistical and systematic, respectively.
$N_{\rm d.o.f}$ is the number of degrees of freedom.\label{tab:FF_combine}}
\scalebox{0.83}{
  \begin{tabular}{l|ccc|ccc|ccc}\hline\hline
     &\multicolumn{3}{c}{Simple pole}  &  \multicolumn{3}{c}{Modified pole}  &  \multicolumn{3}{c}{Series 2-Par} \\
  Decay &  $f_+^{\eta^{(\prime)}}(0)|V_{cs}|$           &  $M_{\rm pole}$        &   $\chi^2/{N_{\rm d.o.f}}$& $f_+^{\eta^{(\prime)}}(0)|V_{cs}|$           &  $\alpha$              &   $\chi^2/{N_{\rm d.o.f}}$& $f_+^{\eta^{(\prime)}}(0)|V_{cs}|$           & $r_1$                  &   $\chi^2/{N_{\rm d.o.f}}$\\\hline
   ${ \eta e^{+} \nu_e}$&$0.4559(61)(63)$    &$1.90(04)(01)$    &13.3/14 &$0.4517(68)(64)$    &  $0.37(08)(03)$  &    12.8/14&$0.4519(71)(65)$    &$-2.63(41)(14)$      &12.8/14\\
   $\eta^{\prime} e^{+} \nu_e$&$0.529(21)(09) $   &$ 1.67(22)(04)$   &0.3/4 &$0.527(23)(09)$    &  $ 0.73(50)(12)$ &    0.2/4 &  $0.525(24)(09)$  & $-6.3(40)(10) $   &0.2/4\\\hline\hline
    \end{tabular}
}
\end{table*}

\section{Summary}

Analyzing 7.33 fb$^{-1}$ $e^+e^-$ collision data taken at center-of-mass energies between $4.128$ GeV and 4.226~GeV with the BESIII detector, the absolute branching fractions of $D^+_s \to\eta e^+ \nu_e$ and
$D^+_s \to\eta^\prime e^+ \nu_e$ are measured.
Compared to Ref.~\cite{bes3_etaev}, which used a subset of the dataset of the present analysis, the precision of the branching fractions of $D^+_s\to \eta e^+\nu_e$ and  $D^+_s\to \eta^\prime e^+\nu_e$ is improved by a factor of 1.3 and 1.7, respectively, and the precision of $f^{\eta^\prime}_+(0)|V_{cs}|$ is improved by a factor of 2.2. 
The precision of $f_+^{\eta}(0)|V_{cs}|$ is not improved because the uncertainty in the previous paper~\cite{bes3_etaev} is underestimated by a factor of two due to incorrect construction of the $\chi^2$ in the fits to the partial decay rates (see ~\cite{correct} for details). 
For simple comparison, we also present the results based on 3.19 fb$^{-1}$ of data at $E_{\rm CM}=4.178$ GeV in the Appendices. After fixing this issue, the precision of $f^{\eta}_+(0)|V_{cs}|$ is improved by a factor of 1.4 as expected. 

\begin{figure*}[htp]\centering
\setlength{\abovecaptionskip}{-2pt}
\setlength{\belowcaptionskip}{-3pt}
\includegraphics[width=0.45\textwidth]{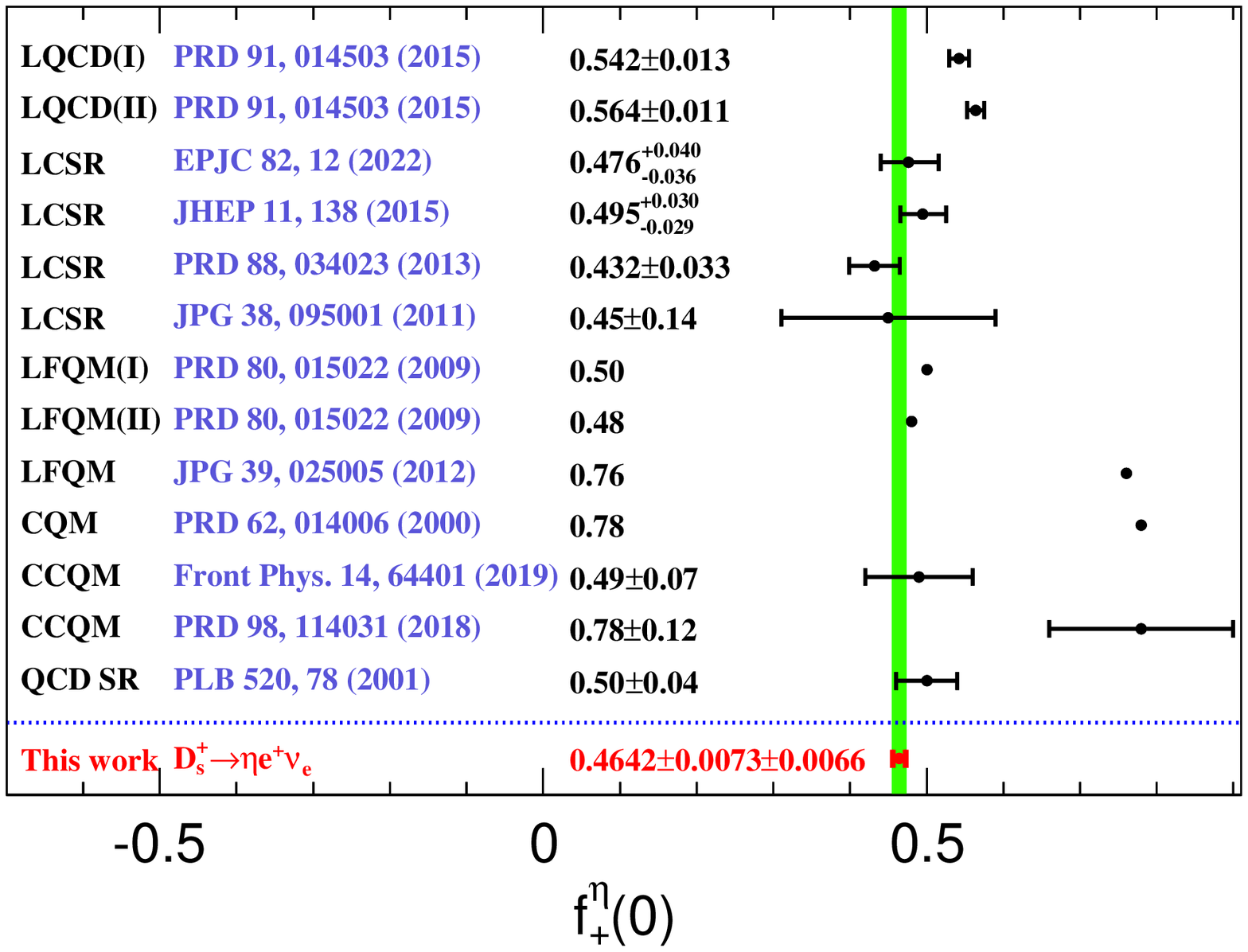}
\includegraphics[width=0.45\textwidth]{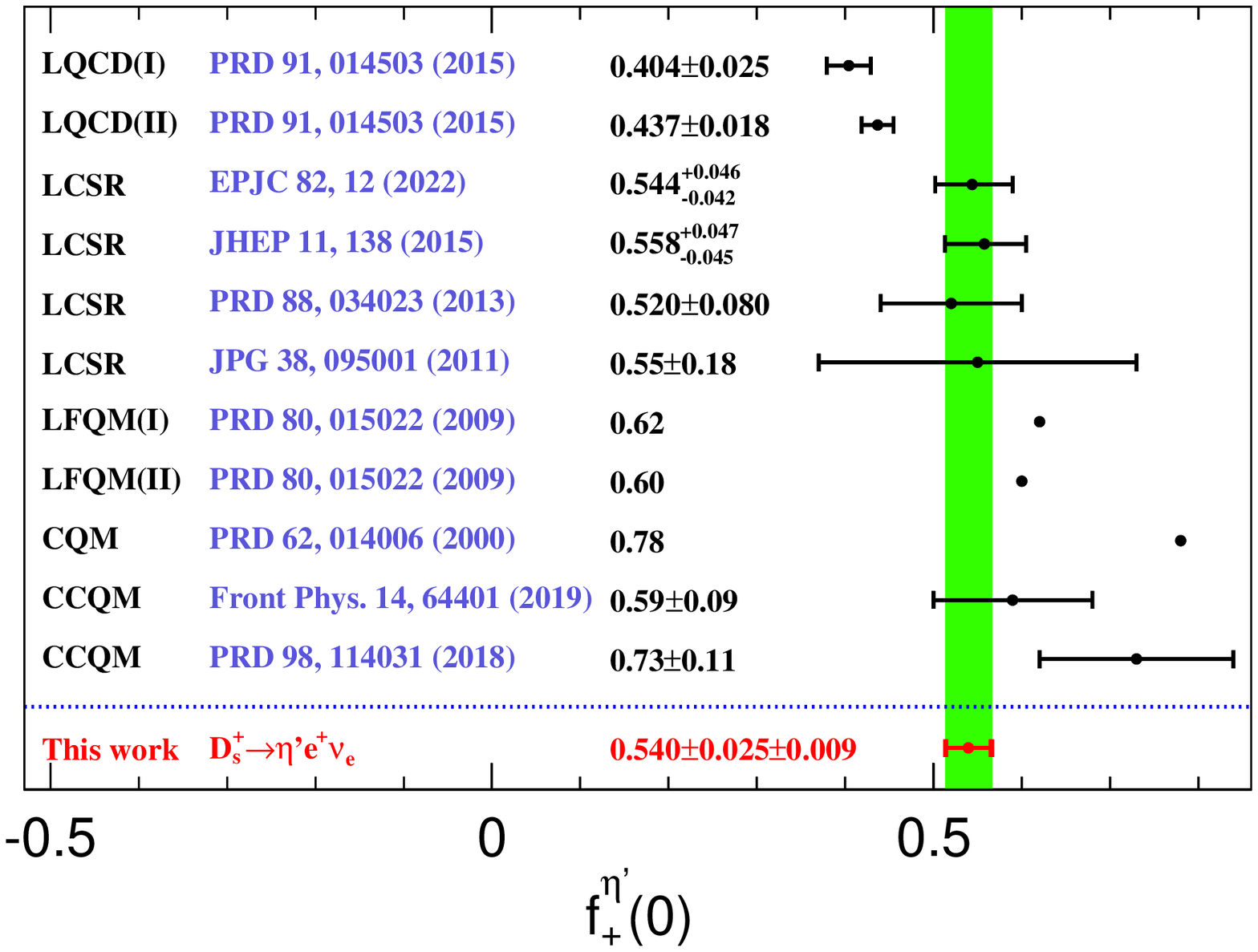}
\caption{Comparisons of the form factors $f_+^{\eta^{(\prime)}}(0)$ measured by this work with the theoretical calculations.
The first and second uncertainties are statistical and systematic, respectively.  The green bands correspond to the $\pm1\sigma$ limits of the form factors measured by this work. For the predictions by LQCD, no systematic uncertainties have been considered. 
}
\label{fig:com_FF}
\end{figure*}

Combining the new branching fractions with those of $D^+ \to\eta e^+ \nu_e$ and
$D^+ \to\eta^\prime e^+ \nu_e$ measured by BESIII~\cite{Zhangyu:Dptoetaenu}, the $\eta-\eta^\prime$ mixing angle $\phi_P=(40.0\pm2.0_{\rm stat}\pm~0.6_{\rm syst})^\circ$ is extracted, providing information related to the gluon component in the $\eta^\prime$ meson. By analyzing the partial decay rates of  $D^+_s \to\eta e^+ \nu_e$ and 
$D^+_s \to\eta^\prime e^+ \nu_e$, the products of $f_+^{\eta^{(\prime)}}(0)|V_{cs}|$ are determined. 
Furthermore, taking the value of $|V_{cs}|$
from a standard model fit (CKM{\sc fitter},~\cite{PDG2022}) as input, 
the form factors at zero momentum transfer squared $f^{\eta^{(\prime)}}_+(0)$ are determined.
The measured hadronic transition form factors provide important pieces of information to test the various theoretical calculations~\cite{Bali:2014pva,Offen:2013nma,Duplancic:2015zna,Melikhov:2000yu,Soni:2018adu,Colangelo:2001cv,Azizi:2010zj}. Figure~\ref{fig:com_FF} shows the comparisons of the $f^{\eta^{(\prime)}}_+(0)$ obtained in this paper and different theoretical predictions.
Alternatively, 
we determine $|V_{cs}|$ with the $D_s^+\to \eta^{(\prime)} e^+\nu_e$ decays by taking the $f_+^{\eta^{(\prime)}}(0)$ given by theoretical calculations~\cite{Duplancic:2015zna}.
 These results on $|V_{cs}|$ together with those
measured by $D\to \bar K\ell^+\nu_\ell$ and $D_s^+\to \ell^+\nu_\ell$ are important to test the
unitarity of the CKM matrix.
The branching fractions, hadronic transition form factors and $|V_{cs}|$ reported in this work supersede the corresponding results in Ref.~\cite{bes3_etaev}, based on the 3.19~fb$^{-1}$ subset of data at $E_{\rm CM}=4.178$~GeV.

%% Saved at => 2022-12-12
\begin{acknowledgments}
The BESIII Collaboration thanks the staff of BEPCII and the IHEP computing center for their strong support. This work is supported in part by National Key R\&D Program of China under Contracts Nos. 2020YFA0406400, 2020YFA0406300; National Natural Science Foundation of China (NSFC) under Contracts Nos. 12305089, 11635010, 11735014, 11835012, 11935015, 11935016, 11935018, 11961141012, 12022510, 12025502, 12035009, 12035013, 12061131003, 12192260, 12192261, 12192262, 12192263, 12192264, 12192265; 
Jiangsu Funding Program for Excellent Postdoctoral Talent under Contracts No. 2023ZB833; 
Project funded by China Postdoctoral Science Foundation under Contracts No. 2023M732547;
the Chinese Academy of Sciences (CAS) Large-Scale Scientific Facility Program; the CAS Center for Excellence in Particle Physics (CCEPP); Joint Large-Scale Scientific Facility Funds of the NSFC and CAS under Contract No. U1932102 and U1832207; CAS Key Research Program of Frontier Sciences under Contracts Nos. QYZDJ-SSW-SLH003, QYZDJ-SSW-SLH040; 100 Talents Program of CAS; The Institute of Nuclear and Particle Physics (INPAC) and Shanghai Key Laboratory for Particle Physics and Cosmology; ERC under Contract No. 758462; European Union's Horizon 2020 research and innovation programme under Marie Sklodowska-Curie grant agreement under Contract No. 894790; German Research Foundation DFG under Contracts Nos. 443159800, 455635585, Collaborative Research Center CRC 1044, FOR5327, GRK 2149; Istituto Nazionale di Fisica Nucleare, Italy; Ministry of Development of Turkey under Contract No. DPT2006K-120470; National Research Foundation of Korea under Contract No. NRF-2022R1A2C1092335; National Science and Technology fund of Mongolia; National Science Research and Innovation Fund (NSRF) via the Program Management Unit for Human Resources \& Institutional Development, Research and Innovation of Thailand under Contract No. B16F640076; Polish National Science Centre under Contract No. 2019/35/O/ST2/02907; The Royal Society, UK under Contract No. DH160214; The Swedish Research Council; U. S. Department of Energy under Contract No. DE-FG02-05ER41374.

\end{acknowledgments}

\clearpage
\appendix
\twocolumngrid
\setcounter{table}{0}
\setcounter{figure}{0}

\section*{Appendices}\label{Supplement}

\section*{\boldmath The results with data taken at $E_{\rm CM}=4.178$ GeV}
\label{App:A}

\begin{figure}[htbp]
\centering
\includegraphics[width=0.48\textwidth]{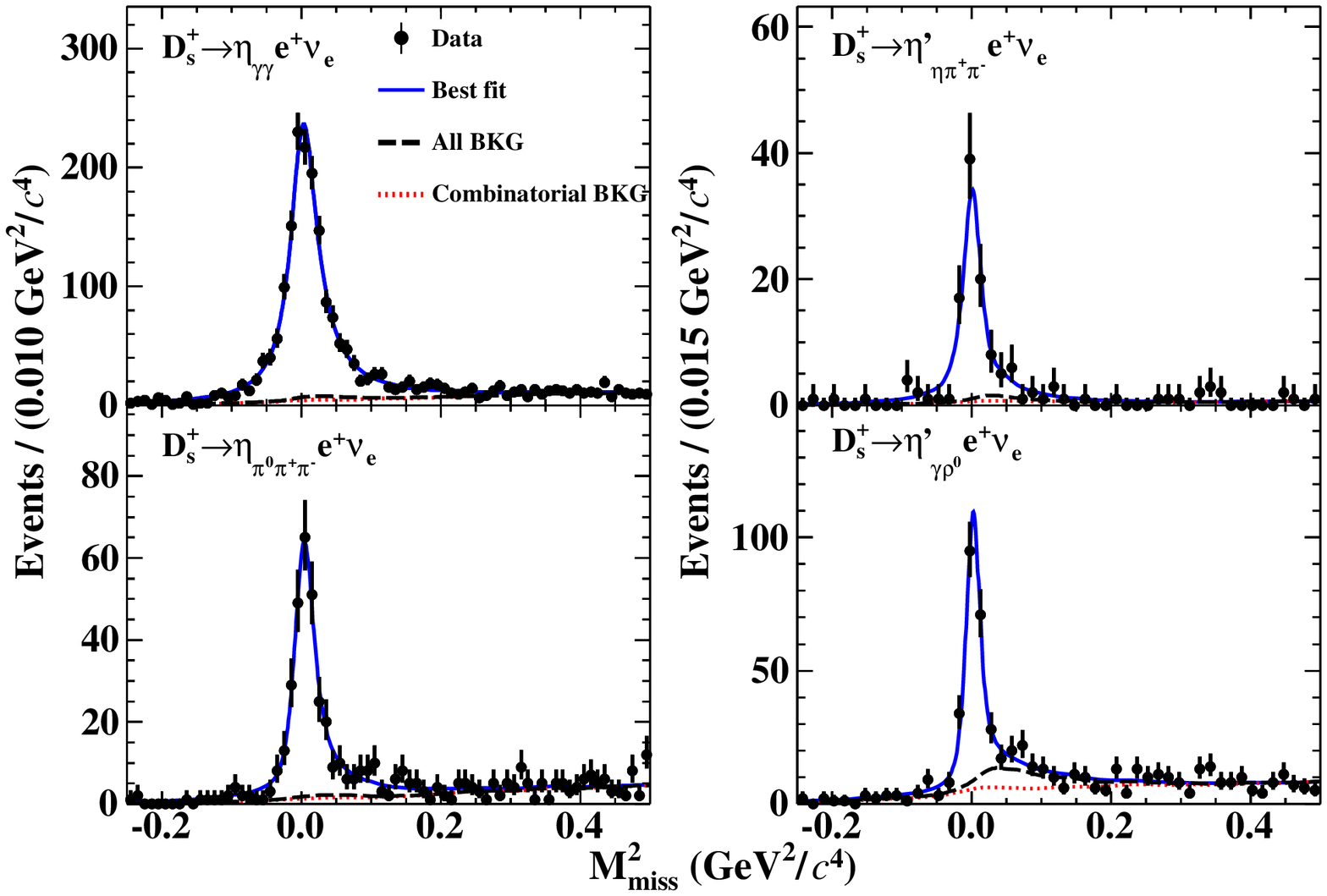}
\caption{Fits to the $M_{\rm miss}^2$ distributions of the candidate events for various semileptonic decays from data taken at $E_{\rm CM}=4.178$ GeV.
The points with error bars represent data. The blue solid curves denote the total fits, and the red solid dotted curves show
the fitted combinatorial background contributions. Differences between dashed and dotted curves are due to the backgrounds from $D_s^+\to\eta^{(\prime)}\pi^+\pi^0$, $\eta^{(\prime)}\mu^+\nu_\mu$, and $D_s^+ \to \phi(1020)_{\pi^0\pi^+\pi^-} e^+ \nu_e$.
\label{fig:4180:fit_ep}}
\end{figure}

Figure~\ref{fig:4180:fit_ep} shows the results of the fits to the $M_{\rm miss}^2$ distributions of the candidate events for $D^+_s\to \eta e^+\nu_e$ and $D^+_s\to \eta^\prime e^+\nu_e$, based on the 3.19 fb$^{-1}$ of $e^+e^-$ collision data taken at $E_{\rm CM}=4.178$ GeV. The obtained signal yields, signal efficiencies and branching fractions
are summarized in Table~\ref{table:par_BF_4180}.

   \begin{table}[hbtp]
   \centering
    \caption{\small Signal efficiencies ($\epsilon_{\rm \gamma(\pi^0){\rm SL}}$), signal yields ($N_{\rm DT}$), and obtained branching fractions ($\mathcal{B}_{\rm SL}$) for various semi-electronic decays  based on the data sample
taken at $E_{\rm CM}=4.178$ GeV.  Efficiencies do not include the branching fractions of $\eta^{(\prime)}$ sub-decays.
Numbers in the first and second parentheses are the statistical and systematic uncertainties, respectively.  \label{table:par_BF_4180}}
\scalebox{0.91}{
\begin{tabular}{lcccc}
\hline\hline
Decay       &$\eta^{(\prime)}$ decays& $\epsilon_{\gamma(\pi^0)\rm SL}$ (\%) & $N_{\rm DT}$                      & $\mathcal B_{\rm SL}$ (\%)  \\ \hline
\multirow{2}{*}{   $\eta e^+\nu_e$ }&    $\gamma\gamma$            & $46.35(11)$         &\multirow{2}{*}{ 2010(49) }&\multirow{2}{*}{2.257(55)(51)}\\
       &  $\pi^0\pi^+\pi^-$      & 17.33(09)         &                                &                                    \\
       \hline
\multirow{2}{*}{$\eta^\prime e^+\nu_e$ }&$\eta\pi^+\pi^-$  & 19.68(07)      &\multirow{2}{*}{337(22) }& \multirow{2}{*}{0.804(53)(22) }\\
    &$\gamma\rho^0$ & 24.26(10)         &                                &                                    \\
\hline\hline
\end{tabular}
}
\end{table}

\begin{figure}[htpb]\centering
\includegraphics[width=0.475\textwidth]{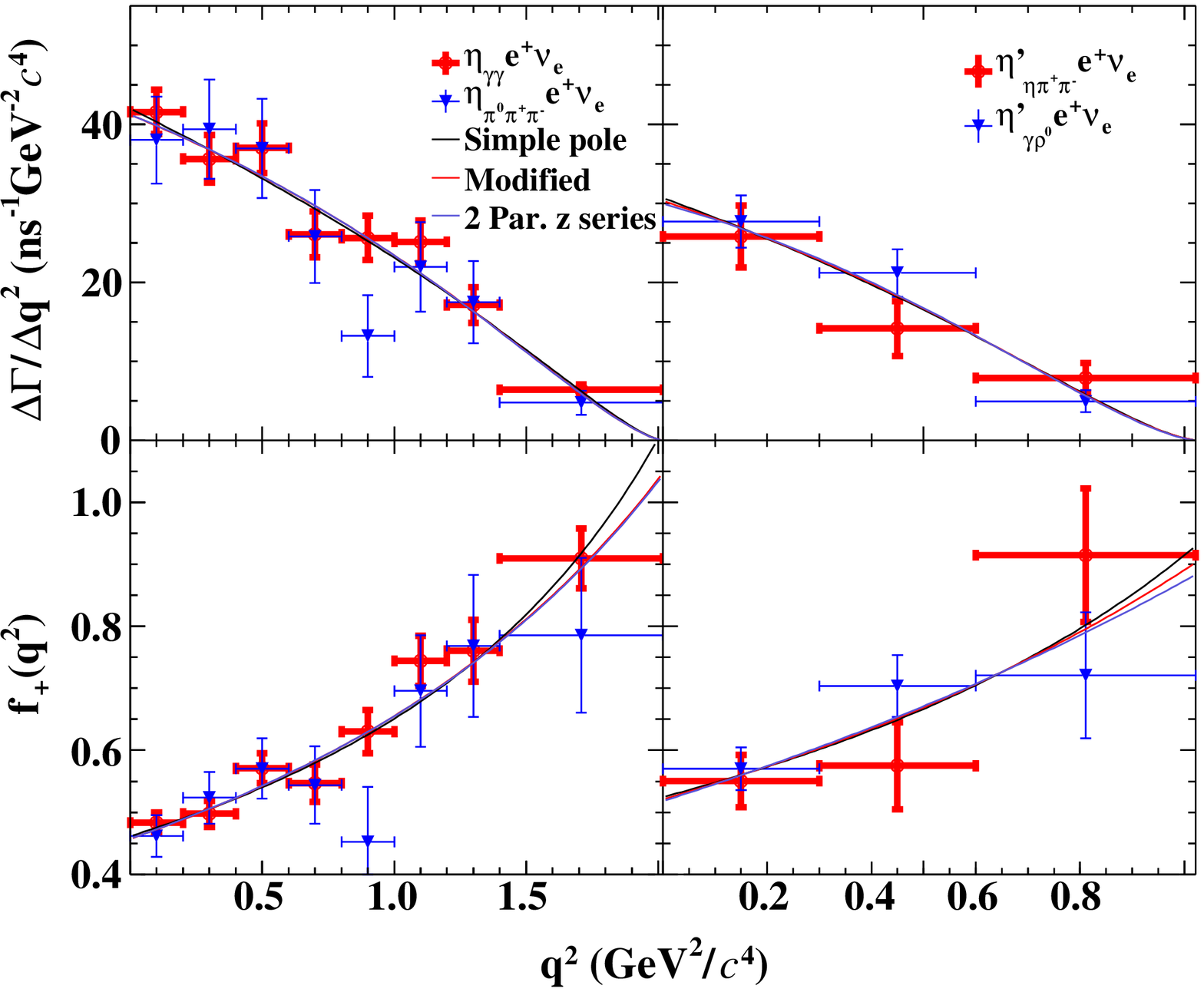}
\caption{
(Top row) Simultaneous fits to the differential decay rates of (left) $D^+_s\to \eta_{\gamma\gamma} e^+\nu_e$ and  $D^+_s\to \eta_{\pi^0\pi^+\pi^-} e^+\nu_e$ and (right) $D^+_s\to \eta^\prime_{\eta\pi^+\pi^-} e^+\nu_e$ and  $D^+_s\to \eta^\prime_{\rho^0\gamma} e^+\nu_e$,
and (bottom row) projection on the hadronic transition form factors as function of $q^2$ for the data sample taken at $E_{\rm CM}=4.178$ GeV. The red circles and blue triangles with error bars are (top row) the measured differential decay rates for two $\eta^{(\prime)}$ channels and (bottom row) the hadronic transition form factors. The black, red, and blue curves are the form factors parameterized by simple pole model, modified pole model, and 2-Par series expansion, respectively.
\label{fig:fitdecayrate_combine4180}}
\end{figure}

Figure~\ref{fig:fitdecayrate_combine4180} shows
the simultaneous fits to the partial decay rates of $D_s^+\to\eta e^+\nu_e$ or $D_s^+\to\eta^\prime e^+\nu_e$ reconstructed with two different decay modes and the hadronic transition form factors as function of $q^2$. The parameters obtained for the hadronic transition form factors are summarized in Table~\ref{tab:FF_combine4180}.

\begin{table*}[htp]
\centering
\caption{\small The parameters obtained from simultaneous fits to the partial decay rates of $D^+_s\to \eta e^+\nu_e$ or
$D^+_s\to \eta^\prime e^+\nu_e$ for  the data sample taken at $E_{\rm CM}=4.178$ GeV.     Numbers in the first and second parentheses are the statistical and systematic uncertainties, respectively.  $N_{\rm d.o.f}$ is the number of degrees of freedom.
\label{tab:FF_combine4180}}
\scalebox{0.83}{
   \begin{tabular}{lccccccccc}\hline\hline
     &\multicolumn{3}{c}{Simple pole}  &  \multicolumn{3}{c}{Modified pole}  &  \multicolumn{3}{c}{Series 2-Par} \\
  Decay &  $f_+^{\eta^{(\prime)}}(0)|V_{cs}|$           &  $M_{\rm pole}$        &   $\chi^2/{N_{\rm d.o.f}}$& $f_+^{\eta^{(\prime)}}(0)|V_{cs}|$           &  $\alpha$              &   $\chi^2/{N_{\rm d.o.f}}$& $f_+^{\eta^{(\prime)}}(0)|V_{cs}|$           & $r_1$                  &   $\chi^2/{N_{\rm d.o.f}}$\\\hline
   ${ \eta e^{+} \nu_e}$&$0.4494(085)(067)$    &$1.85(05)(02)$    &13.6/14 &$0.4452(095)(068)$    &  $0.45(11)(04)$  &    13.9/14&$0.4454(101)(068)$    &$-2.99(58)(18)$      &13.9/14\\
   $\eta^{\prime} e^{+} \nu_e$&$0.511(28)(08) $   &$ 1.53(19)(02)$   &3.8/4 &$0.507(30)(08)$    &  $ 1.10(59)(07)$ &    3.8/4 &  $0.504(34)(09)$  & $-9.3(50)(06) $   &3.8/4\\\hline\hline
    \end{tabular}

}
\end{table*}

Tables \ref{tab:4180:decayratea} and \ref{tab:4180:decayrateb}
summarize the $q^{2}$ ranges, the fitted numbers of observed events ($N_{\rm DT}$),
the numbers of generated events ($N_{\rm prd}$) calculated by the weighted efficiency matrices
and the decay rates ($\Delta\Gamma$) of $D_s^+\to\eta e^+\nu_e$ and $D_s^+\to\eta^\prime e^+\nu_e$ in various $q^{2}$ intervals, respectively. 

Tables \ref{tab:4180:cova} and \ref{tab:4180:covb}
show the statistical and systematic uncertainty covariance density matrices for $D_s^+\to\eta e^+\nu_e$ and $D_s^+\to\eta^\prime e^+\nu_e$, respectively.

\begin{table*}[htp]\centering
\caption{The partial decay rates of $D_s^+\to\eta e^+\nu_e$ in various $q^{2}$ intervals of data for the data sample taken at $E_{\rm CM}=4.178$ GeV. Numbers in the parentheses are the statistical uncertainties. }
\label{tab:4180:decayratea}
\scalebox{0.85}{
\begin{tabular}{cccccccccc}\hline\hline
\multicolumn{2}{c}{$i$}&1&2&3&4&5&6&7&8\\
\multicolumn{2}{c}{$q^2$ $(\mathrm{GeV}^{2}/c^{4})$}&($m_e^2,\,0.2$)&($0.2,\,0.4$)&($0.4,\,0.6$)&($0.6,\,0.8$)&($0.8,\,1.0$)&($1.0,\,1.2$)&($1.2,\,1.4$)&($1.4,\,2.02$)\\
\hline
\multirow{3}{*}{$D_s^+\to\eta_{\gamma\gamma} e^+\nu_e$}&$N_{\mathrm{DT}}^i$&320(19)&274(18)&264(18)&199(16)&185(15)&173(14)&122(12)&142(14)\\
&$N_{\mathrm{prd}}^i$&1682(113)&1441(122)&1497(127)&1055(117)&1037(114)&1016(110)&692(091)&797(085)\\
&$\Delta\Gamma^i_{\rm msr}$ $(\mathrm{ns^{-1}})$&8.32(56)&7.13(61)&7.40(63)&5.21(58)&5.12(56)&5.02(54)&3.42(45)&3.94(42)\\
\hline
\multirow{3}{*}{$D_s^+\to\eta_{\pi^0\pi^+\pi^-} e^+\nu_e$}&$N_{\mathrm{DT}}^i$&70(09)&68(09)&59(08)&40(07)&23(06)&29(06)&23(05)&20(05)\\
&$N_{\mathrm{prd}}^i$&1538(222)&1594(255)&1495(256)&1044(238)&534(210)&888(229)&707(210)&594(189)\\
&$\Delta\Gamma^i_{\rm msr}$ $(\mathrm{ns^{-1}})$&7.60(110)&7.88(126)&7.39(126)&5.16(118)&2.64(104)&4.39(113)&3.49(104)&2.94(093)\\\hline\hline
\end{tabular}
}
\end{table*}

\begin{table*}[htp]\centering
\caption{The partial decay rates of $D_s^+\to\eta^\prime e^+\nu_e$ in various $q^{2}$ intervals of data for the data sample taken at $E_{\rm CM}=4.178$ GeV. Numbers in the parentheses are the statistical uncertainties. }
\label{tab:4180:decayrateb}
\scalebox{0.9}{
\begin{tabular}{ccccc}\hline\hline
\multicolumn{2}{c}{$i$}&1&2&3\\
\multicolumn{2}{c}{$q^2$ $(\mathrm{GeV}^{2}/c^{4})$}&($m_e^2,\,0.3$)&($0.3,\,0.6$)&($0.6,\,1.02$)\\
\hline
\multirow{3}{*}{$D_s^+\to\eta^\prime_{\eta_{\gamma\gamma}\pi^+\pi^-} e^+\nu_e$}&$N_{\mathrm{DT}}^i$&52(08)&29(06)&21(05)\\
&$N_{\mathrm{prd}}^i$&1566(238)&858(210)&673(159)\\
&$\Delta\Gamma^i_{\rm msr}$ $(\mathrm{ns^{-1}})$&7.7(12)&4.2(10)&3.3(08)\\
\hline
\multirow{3}{*}{$D_s^+\to\eta^\prime_{\gamma\rho^0} e^+\nu_e$}&$N_{\mathrm{DT}}^i$&120(14)&90(12)&32(08)\\
&$N_{\mathrm{prd}}^i$&1680(202)&1285(182)&418(118)\\
&$\Delta\Gamma^i_{\rm msr}$ $(\mathrm{ns^{-1}})$&8.3(10)&6.4(09)&2.1(06)\\
\hline\hline
\end{tabular}
}
\end{table*}

\begin{table*}[htp]\centering
\caption{Statistical and systematic uncertainty density matrices for the measured partial decay rates of $D_s^+\to\eta e^+\nu_e$ in different $q^2$ intervals.}
\label{tab:4180:cova}
\scalebox{0.9}{
\begin{tabular}{ccccccccccccccccc}\hline\hline
\multicolumn{17}{c}{Statistical correlation matrix}\\
\multirow{2}{*}{$\rho_{ij}^{\rm stat}$}&\multicolumn{8}{c}{$D_s^+\to\eta_{\gamma\gamma}e^+\nu_e$}&\multicolumn{8}{c}{$D_s^+\to\eta_{\pi^0\pi^+\pi^-}e^+\nu_e$}\\
&1&2&3&4&5&6&7&8&1&2&3&4&5&6&7&8\\
\hline
1	&1.000	&-0.181	&0.017	&-0.003	&-0.001	&0.000	&-0.001	&0.000	&0.000	&0.000	&0.000	&0.000	&0.000	&0.000&0.000	&0.000	\\
2	&&1.000	&-0.232	&0.025	&-0.004	&0.000	&0.000	&-0.002	&0.000	&0.000	&0.000	&0.000	&0.000	&0.000	&0.000&0.000	\\
3	&&&1.000	&-0.248	&0.024	&-0.005	&0.000	&0.000	&0.000	&0.000	&0.000	&0.000	&0.000	&0.000	&0.000&0.000	\\
4	&&&&1.000	&-0.252	&0.026	&-0.005	&-0.001	&0.000	&0.000	&0.000	&0.000	&0.000	&0.000	&0.000	&0.000\\
5	&&&&&1.000	&-0.247	&0.027	&-0.005	&0.000	&0.000	&0.000	&0.000	&0.000	&0.000	&0.000	&0.000	\\
6	&&&&&&1.000	&-0.239	&0.012	&0.000	&0.000	&0.000	&0.000	&0.000	&0.000	&0.000	&0.000	\\
7	&&&&&&&1.000	&-0.159	&0.000	&0.000	&0.000	&0.000	&0.000	&0.000	&0.000	&0.000	\\
8	&&&&&&&&1.000	&0.000	&0.000	&0.000	&0.000	&0.000	&0.000	&0.000	&0.000	\\
1	&&&&&&&&&1.000	&-0.167	&0.011	&-0.001	&0.000	&-0.001	&0.000	&0.000	\\
2	&&&&&&&&&&1.000	&-0.207	&0.007	&-0.004	&-0.001	&-0.001	&0.000	\\
3	&&&&&&&&&&&1.000	&-0.208	&0.010	&0.001	&-0.001	&-0.001	\\
4	&&&&&&&&&&&&1.000	&-0.225	&0.008	&-0.003	&0.000	\\
5	&&&&&&&&&&&&&1.000	&-0.226	&0.016	&-0.002	\\
6	&&&&&&&&&&&&&&1.000	&-0.231	&0.008	\\
7	&&&&&&&&&&&&&&&1.000	&-0.183	\\
8	&&&&&&&&&&&&&&&&1.000	\\

\hline
\multicolumn{17}{c}{Systematic correlation matrix}\\
\multirow{2}{*}{$\rho_{ij}^{\rm syst}$}&\multicolumn{8}{c}{$D_s^+\to\eta_{\gamma\gamma}e^+\nu_e$}&\multicolumn{8}{c}{$D_s^+\to\eta_{\pi^0\pi^+\pi^-}e^+\nu_e$}\\
&1&2&3&4&5&6&7&8&1&2&3&4&5&6&7&8\\
\hline
1	&1.000	&0.504	&-0.024	&0.461	&0.715	&0.834	&0.777	&0.821	&0.376	&0.374	&-0.189	&0.618	&-0.243	&0.323&-0.123	&0.551	\\
2	&&1.000	&0.769	&0.621	&0.709	&0.520	&0.207	&0.182	&0.546	&0.534	&0.633	&-0.158	&0.593	&0.580	&0.607&-0.324	\\
3	&&&1.000	&0.412	&0.358	&0.060	&-0.288	&-0.362	&0.406	&0.393	&0.905	&-0.622	&0.893	&0.486	&0.827&-0.778	\\
4	&&&&1.000	&0.429	&0.551	&0.520	&0.307	&0.359	&0.351	&0.305	&0.020	&0.273	&0.357	&0.297	&-0.083\\
5	&&&&&1.000	&0.656	&0.479	&0.573	&0.480	&0.470	&0.215	&0.248	&0.162	&0.446	&0.225	&0.123	\\
6	&&&&&&1.000	&0.713	&0.755	&0.381	&0.374	&-0.099	&0.506	&-0.153	&0.313	&-0.059	&0.425	\\
7	&&&&&&&1.000	&0.802	&0.197	&0.196	&-0.412	&0.677	&-0.456	&0.110	&-0.347	&0.661	\\
8	&&&&&&&&1.000	&0.206	&0.205	&-0.473	&0.757	&-0.521	&0.109	&-0.400	&0.743	\\
1	&&&&&&&&&1.000	&0.903	&0.568	&0.318	&0.497	&0.884	&0.604	&0.064	\\
2	&&&&&&&&&&1.000	&0.544	&0.313	&0.475	&0.870	&0.588	&0.073	\\
3	&&&&&&&&&&&1.000	&-0.552	&0.959	&0.629	&0.913	&-0.746	\\
4	&&&&&&&&&&&&1.000	&-0.589	&0.200	&-0.408	&0.939	\\
5	&&&&&&&&&&&&&1.000	&0.558	&0.942	&-0.755	\\
6	&&&&&&&&&&&&&&1.000	&0.651	&-0.039	\\
7	&&&&&&&&&&&&&&&1.000	&-0.597	\\
8	&&&&&&&&&&&&&&&&1.000	\\

\hline\hline
\end{tabular}
}
\end{table*}

\begin{table*}[htp]\centering
\caption{Statistical and systematic uncertainty density matrices for the measured partial decay rates of  $D_s^+\to\eta^\prime e^+\nu_e$ in different
$q^2$ intervals.}
\label{tab:4180:covb}
\begin{tabular}{ccccccc}\hline\hline
\multicolumn{7}{c}{Statistical correlation matrix}\\
\multirow{2}{*}{$\rho_{ij}^{\rm stat}$}&\multicolumn{3}{c}{$D_s^+\to\eta^\prime_{\eta\pi^+\pi^-}e^+\nu_e$}&\multicolumn{3}{c}{$D_s^+\to\eta^\prime_{\gamma\rho^0}e^+\nu_e$}\\
&1&2&3&1&2&3\\
\hline
1	&1.000	&-0.119	&0.008	&0.000	&0.000	&0.000	\\
2	&&1.000	&-0.124	&0.000	&0.000	&0.000	\\
3	&&&1.000	&0.000	&0.000	&0.000	\\
1	&&&&1.000	&-0.103	&0.003	\\
2	&&&&&1.000	&-0.102	\\
3	&&&&&&1.000	\\\hline
\multicolumn{7}{c}{Systematic correlation matrix}\\
\multirow{2}{*}{$\rho_{ij}^{\rm syst}$}&\multicolumn{3}{c}{$D_s^+\to\eta^\prime_{\eta\pi^+\pi^-}e^+\nu_e$}&\multicolumn{3}{c}{$D_s^+\to\eta^\prime_{\gamma\rho^0}e^+\nu_e$}\\
&1&2&3&1&2&3\\
\hline
1	&1.000	&0.678	&0.822	&0.222	&0.354	&0.280	\\
2	&&1.000	&0.946	&0.485	&0.134	&0.093	\\
3	&&&1.000	&0.429	&0.226	&0.170	\\
1	&&&&1.000	&0.692	&0.549	\\
2	&&&&&1.000	&0.744	\\
3	&&&&&&1.000	\\
\hline\hline
\end{tabular}
\end{table*}


\begin{thebibliography}{30}
\bibitem{correct}
In Ref.~\cite{bes3_etaev}, the $\chi^2$ is constructed to be  $\chi^{2} = \sum\limits_{ij}\left(\Delta\Gamma_{i}^{\mathrm{msr}}-\Delta\Gamma_{i}^{\mathrm{th}}\right) C_{ij}^{-1}\left(\Delta\Gamma_{j}^{\mathrm{msr}}-\Delta\Gamma_{j}^{\mathrm{th}}\right)/\sigma(\Delta\Gamma_i^{\rm msr})/\sigma(\Delta\Gamma_j^{\rm msr})$, where there is an additional term of $\sigma(\Delta\Gamma^{\rm msr}_i)\sigma(\Delta\Gamma^{\rm msr}_j)$ used for test but not removed when obtaining nominal results. This mistake only changes $f^{\eta}_+(0)|V_{cs}|$ by 0.2\%, but underestimate its uncertainty by a factor of 2.
\end{thebibliography}
\end{document}